\documentclass[twocolumn,tighten,times]{aastex701}
 
\usepackage{mathptmx}
\usepackage[T1]{fontenc}
\usepackage{textcomp}

\journalinfo{Accepted for publication in ApJ | Current Version Date: \today}
\usepackage{amssymb,amsmath}
\usepackage{bm}
\usepackage{multirow}
\usepackage{nicefrac}
\usepackage{array}
\usepackage{booktabs}
\graphicspath{{./}{./figures/}}
\shortauthors{Schad et al.}
\shorttitle{Near-Sun Sky Diagnostics at Mauna Loa}

\newcommand{\okina}{\textquoteleft}
\usepackage{hyperref}

\begin{document}

\title{\Large{\textbf{Joint Diagnostics of Circumsolar Sky Brightness Using Coronagraphic \\ Measurements and Aerosol Optical Inversions at Mauna Loa}}}

\author[0000-0002-7451-9804,sname="Schad"]{Thomas A. Schad}
\email[show]{tschad@nso.edu}
\correspondingauthor{Thomas A. Schad}
\affiliation{National Solar Observatory, 22 \okina Ohi\okina a K\={u} Street, Pukalani, HI 96768, USA}

% -------------------------------------------------------------------
%  Co-authors alphabetical by last name
% -------------------------------------------------------------------

\author[0000-0001-5681-9689,sname="Bryans"]{Paul Bryans}
\email{pbryans@ucar.edu}
\affiliation{High Altitude Observatory, National Center for Atmospheric Research, Boulder, CO 80301, USA}

\author[0000-0002-7978-368X,sname="Fehlmann"]{Andre Fehlmann}
\email{afehlmann@nso.edu}
\affiliation{National Solar Observatory, 22 \okina Ohi\okina a K\={u} Street, Pukalani, HI 96768, USA}

\author[0000-0001-9831-2640,sname="Gibson"]{Sarah Gibson}
\email{sgibson@ucar.edu}
\affiliation{High Altitude Observatory, National Center for Atmospheric Research, Boulder, CO 80301, USA}

\author[0000-0002-3215-7155,sname="Harrington"]{David M. Harrington}
\email{dharrington@nso.edu}
\affiliation{National Solar Observatory, 22 \okina Ohi\okina a K\={u} Street, Pukalani, HI 96768, USA}

\author[0000-0002-8259-8303,sname="Tarr"]{Lucas A. Tarr}
\email{ltarr@nso.edu}
\affiliation{National Solar Observatory, 22 \okina Ohi\okina a K\={u} Street, Pukalani, HI 96768, USA}

\author[0000-0001-7399-3013,sname="Tomczyk"]{Steven Tomczyk}
\email{steventomczyk23@gmail.com}
\affiliation{Solar Scientific LLC, Boulder, CO 80301, USA}

\author[0009-0004-0302-1218,sname="Yepez"]{Jeffrey G. Yepez}
\email{jgyepez@hawaii.edu}
\affiliation{Akamai Workforce Initiative, Institute for Scientist \& Engineer Educators, University of California Observatories, Santa Cruz, CA 95064, USA}
\affiliation{Department of Physics and Astronomy, University of Hawai\okina i at M\=anoa, Honolulu, HI 96822, USA}

%\author[sname="Yim"]{Kevin Yim}
%\email{kpy9@case.edu}
%\affiliation{Akamai Workforce Initiative, Institute for Scientist \& Engineer Educators, University of California Observatories, Santa Cruz, CA 95064, USA}
%\affiliation{Department of Physics, Case Western Reserve University, Cleveland, OH 44106, USA}

%%%%%%%%%%%%%%%%%%%%%%%%%%%%%%%%%%%%%%%%%%%%%%%%%%%%%%%%%%%%%%%%%%%%
\begin{abstract}
\noindent Atmospheric aerosols strongly influence daytime sky quality for solar coronal imaging, yet few studies directly link aerosol properties and sky brightness measurements within $\sim\!2^{\circ}$ of the Sun. Here, we compare externally occulted coronagraphic measurements of near-Sun radiance with aerosol-constrained inferences derived from direct-Sun and sky photometry. Our analysis focuses on Mauna Loa Observatory, a well-characterized high-altitude site for atmospheric and solar observations. We present coronagraphic measurements of near-Sun radiance at $1.54 \pm 0.77^\circ$ from solar disk center acquired between 2006 and 2007 by an ATST Sky Brightness Monitor (SBM). These data are directly compared with circumsolar radiances inferred at $1.54^{\circ}$ using AERONET almucantar measurements and aerosol optical retrievals. We find quantitative agreement between these two approaches, enabling extension to multi-decadal analyses of circumsolar radiance and its relationship to aerosol properties and related proxies (\textit{e.g.}, the Ångström exponent) using AERONET data from 2000--2025.  Near-Sun radiances are expressed relative to the solar disk-center radiance, facilitating direct comparison to related studies. Finally, we synthesize physically based true-color images of the circumsolar sky under representative aerosol conditions as an observational aid, in part to illustrate that visually enhanced solar aureoles do not necessarily imply poor infrared coronal observing conditions. This methodology provides an extended framework for assessing daytime coronal sky quality at existing and future observing sites, including DKIST and the proposed COSMO facility.
\end{abstract}


%%%%%%%%%%%%%%%%%%%%%%%%%%%%%%%%%%%%%%%%%%%%%%%%%%%%%%%%%%%%%%%%%%%%
%%% SECTION 1 -- INTRODUCTION
%%%%%%%%%%%%%%%%%%%%%%%%%%%%%%%%%%%%%%%%%%%%%%%%%%%%%%%%%%%%%%%%%%%%

\section{Introduction} \label{sec:intro}
\setcounter{footnote}{0}  % next \footnote will be 1

The near-Sun daytime sky brightness largely governs the observability of the solar corona from ground-based telescopes. Outside total solar eclipses, atmospheric scattering of solar-disk light produces a sky background that overwhelms the coronal signal at visible and infrared wavelengths.  The peak intensity of the off-limb coronal K-continuum \citep{schuster1879, minnaert1930} is only $\sim\!1$ part per million of the mean solar-disk radiance, or $\mu B_{\odot}$, and decreases rapidly with increasing elongation. Meanwhile, spectrally-resolved coronal emission lines can reach intensities of order tens of $\mu B_{\odot}$ near the solar limb but also produce weak yet diagnostically-important fractional polarization signatures extending down to $\lesssim 10^{-4}$ \citep{lin2000,schad2024SciA}.  Thus, coronal observations are background-limited \citep{penn2004} and benefit substantially from the careful selection of observing locations in addition to low-stray-light coronagraphic instrumentation \citep{lyot1939}.

Under cloud-free conditions, the near-Sun sky radiance arises from sunlight scattered by molecules (Rayleigh) and aerosols (particle scattering), with aerosols often providing the larger contribution.  Across atmospheric science, meteorology, and solar energy research \citep[e.g.,][]{abreu2020}, this near-Sun region is variously referred to as circumsolar radiance, the sunshape function, or the solar aureole, with an angular extent frequently referenced to the full-angle field-of-view (FOV) of normal-incidence pyrheliometers ($\sim\!5$--$5.7^{\circ}$; half-angle $\sim\!2.5$--$2.85^{\circ}$). The angular and spectral structure of the circumsolar radiance is sensitive to aerosol microphysics, especially particle size, and these same aerosol microphysical properties also determine atmospheric turbidity and influence the partitioning of direct-normal, diffuse, and global solar radiation. 

Although directly relevant, atmospheric composition and air quality studies rarely include direct measurements of the near-Sun sky radiance within $\lesssim\!2^{\circ}$ of solar disk center \citep{abreu2020}; instead they rely on spaceborne and field-based observations (e.g., direct-Sun spectrophotometry, sky radiometry, satellite-based spectroradiometry, lidar, and standardized in situ measurements) that retrieve quantities physically linked to circumsolar scattering.  This makes them diagnostically valuable for longer term assessment of daytime sky quality provided the near-Sun radiance can be properly inferred.

For solar coronal observations, both the angular and spectral distributions of near-Sun sky radiance are important, especially within $\sim\!1^\circ$ from solar disk center, \textit{i.e.} $\lesssim 4\,R_{\odot}$\footnote{The apparent solar angular radius as viewed from Earth varies between $0.262^{\circ}$ to $0.271^{\circ}$ during the year.}. Achieving sensitivity better than 1 $\mu B_{\odot}$ in this region requires advanced coronagraphic techniques, which can be impractical for routine, multi-site atmospheric monitoring. Early efforts to help assess coronal sky quality include the novel visual sky photometers developed by \citet{evans1948} to measure the radiance between 1.6 and $4.4\,R_{\odot}$ (\textit{i.e.} $\sim\!0.5$ to $1.2^{\circ}$) at 530 nm.  These devices were deployed in multiple locations, including at Haleakalā \citep{LaBonte2003, Lin2004PASP} and for short deployments at Mauna Loa \citep{garcia1983}.  In support of the site survey for the Advanced Technology Solar Telescope \citep{Hill2006}, now named the U.S. NSF Daniel K. Inouye Solar Telescope \citep{rimmele2020}, a multi-wavelength imaging-based Sky Brightness Monitor (SBM) was developed by \citet{Lin2004PASP} to observe the near-Sun region below $2^{\circ}$.  Nevertheless, these techniques are rarely coupled with detailed atmospheric composition measurements, limiting insight into the microphysical origins of favorable coronal observing sites and their long-term variability.

In this article, we examine the linkage between aerosol optical properties and circumsolar radiance at Mauna Loa Observatory. Section~\ref{sec:data} describes the observational data sets used in this study, including coronagraphic sky-brightness measurements, sun--sky photometry, and broadband radiation observations. Section~\ref{sec:daily_comparisons} establishes the empirical connection between these data through detailed daily case studies spanning distinct aerosol regimes. Building on these comparisons, Section~\ref{sec:aeronet_model} develops a simplified forward model that uses AERONET-retrieved aerosol properties to infer the near-Sun sky radiance closer to the Sun than it directly observes. Section~\ref{sec:statistics} then evaluates the statistical consistency between the AERONET inferred near-Sun radiances and the SBM data, as well as their relationship to other measures of atmospheric conditions.  Section~\ref{sec:temporal_variability} analyzes the climatology of the inferred circumsolar brightness across seasonal and diurnal timescales. Finally, Section~\ref{sec:synthColor} applies a more comprehensive radiative transfer framework to synthesize true-color images of the circumsolar sky under representative aerosol conditions, providing an observer-oriented visualization of these effects.

%%%%%%%%%%%%%%%%%%%%%%%%%%%%%%%%%%%%%%%%%%%%%%%%%%%%%%%%%%%%%%%%%%%%
%%% SECTION 2
%%%%%%%%%%%%%%%%%%%%%%%%%%%%%%%%%%%%%%%%%%%%%%%%%%%%%%%%%%%%%%%%%%%%

\begin{deluxetable*}{llll}
\tabletypesize{\footnotesize}
\tablecaption{Summary of remote-sensing measurements used to characterize daytime circumsolar conditions at Mauna Loa. Columns list the data source, sampling time interval, spectral passbands or radiometric units, and representative uncertainties.}
\label{tbl:data_sources}
\tablehead{
  \colhead{Parameter} & \colhead{Time Frame} & \colhead{Bands / Units} & \colhead{Notes / Typical Uncertainty}
}
\startdata
\multicolumn{4}{c}{\textbf{ATST Sky Brightness Monitor \citep[SBM;][]{Lin2004PASP}}} \\
\hline
Disk-center radiance & 2006 Jun 11 -- 2007 Jun 21 & 450/530/890/940\,nm & Arbitrary radiance units \\
Mean near-Sun radiance & 2006 Jun 11 -- 2007 Jun 21 & $4.5$--$7.8\,R_{\odot}$ annulus & Relative to disk radiance; relative uncertainty $\lesssim 10\%$\tablenotemark{a} \\
\\[-4pt]
\multicolumn{4}{c}{\textbf{AERONET Version~3 \citep{Holben1998,Dubovik2000Accuracy,Giles2019}}} \\
\hline
Aerosol optical depth (AOD) & 2000--present\tablenotemark{b} & 440/675/870/1020\,nm & Direct-Sun retrieval; uncertainty $\sim\!0.01$--$0.02$\tablenotemark{c} \\
Ångström exponent, $\alpha$ & 2000--present\tablenotemark{b} & dimensionless & Derived from AOD spectral slope \\
Almucantar radiances & 2000--present\tablenotemark{b} & radiance & For inversion; relative uncertainty $\sim\!5\%$\tablenotemark{c} \\
Aerosol size distribution & 2000--present\tablenotemark{b} & $dV/d\ln(r)$ & Inversion-derived; typically $\lesssim 10\%$\tablenotemark{c} \\
Aerosol phase function & 2000--present\tablenotemark{b} & dimensionless & Normalized to $4\pi$; typically $\sim\!5\%$\tablenotemark{c} \\
\\[-4pt]
\multicolumn{4}{c}{\textbf{NOAA/GML Baseline Radiation Measurements (MLO)}} \\
\hline
Direct-normal irradiance & 2001--present & $\mathrm{W\,m^{-2}}$ (280--3000\,nm) & Pyrheliometer; calibration accuracy 1--2\% \\
Diffuse irradiance & 2001--present & $\mathrm{W\,m^{-2}}$ (280--3000\,nm) & Pyranometer; typical uncertainty 3--5\% \\
\enddata
\tablecomments{%
\tablenotetext{a}{Estimated from published SBM stability and cross-comparison results \citep{Lin2004PASP}.}%
\vspace{-0.45\baselineskip}%
\tablenotetext{b}{Although AERONET observations at Mauna Loa began in 1994, this study restricts analysis to data acquired from 2000 onward, corresponding to adoption of a consistent operational configuration.}%
\vspace{-0.45\baselineskip}%
\tablenotetext{c}{AERONET Version~3 uncertainties: direct-Sun AOD $\sim\!0.01$--$0.02$; almucantar radiances $\sim\!5\%$; size distributions and phase functions typically $\lesssim 10\%$ \citep{Dubovik2000Accuracy,Giles2019,Sinyuk2020}.}%
}
\end{deluxetable*}

\section{Observational Data and Methods} \label{sec:data} 

\subsection{Site Overview: Mauna Loa Observatory}\label{sec:site}

This study focuses on measurements acquired at the Mauna Loa Observatory (MLO), located at an elevation of 3397~m above sea level on the Island of Hawai\okina i ($19.536^\circ$~N, $155.576^\circ$~W). MLO’s high altitude, remote location in the central Pacific, and limited influence from local anthropogenic sources make it a premier site for both solar observations and long-term atmospheric monitoring. The observatory is frequently situated above the marine boundary layer, enabling routine sampling of free-tropospheric air masses under cloud-free conditions \citep[e.g.,][]{Price1963, dutton2001, park2020}.

From a solar observing perspective, MLO has long been recognized for its favorable daytime sky conditions, particularly for coronal observations \citep{Fischer1981}. From an atmospheric science perspective, MLO serves as one of NOAA’s baseline observatories and as a primary calibration site for AERONET, providing a uniquely well-characterized, multi-decadal record of aerosol optical properties \citep{Holben1998, dutton2001}. These complementary roles make MLO an ideal testbed for examining the linkage between aerosol microphysics and circumsolar sky brightness.

Importantly, MLO is used here not as a representative global site.  It generally has very low aerosol loading; however, it also experiences episodic influences from long-range particulate transport (e.g., Asian dust and pollution), which enables the study of circumsolar scattering across a wider dynamic range. Demonstrating agreement between coronagraphic measurements and aerosol-constrained inferences under these conditions therefore provides a valid test of the methodology developed in this work. An overview of the data sources, temporal coverage, spectral bands, and representative uncertainties is provided in Table~\ref{tbl:data_sources}.

\subsection{Coronagraphic Measurements of the Near-Sun Sky Radiance (SBM)} 

Direct image-based measurements of the near-Sun sky brightness were obtained using an ATST Sky Brightness Monitor \citep[SBM;][]{Lin2004PASP}, temporarily deployed on the guided spar of the Mauna Loa Solar Observatory\footnote{\url{https://www2.hao.ucar.edu/mlso}} between 11 June 2006 and 21 June 2007. The SBM is an externally occulted telescope capable of imaging the solar-disk and the surrounding corona simultaneously on a single detector.  By attenuating the solar-disk with a calibrated neutral-density filter, the instrument achieves a relative sensitivity for sky measurements better than $10^{-6}\,B_{\odot}$. During operation, the SBM measures the mean near-Sun sky radiance within an FOV spanning elongations approximately $1.12$--$1.95^{\circ}$ from Sun center ($\sim\!4.5$ to $7.8\,R_{\odot}$) at the same time as the solar disk radiance. The instrument observes in four 10-nm-wide spectral bandpasses centered at 450, 530, 890, and 940~nm.  The 940~nm channel is sensitive to the precipitable water vapor.

SBM raw observations consist of five 30~s exposures acquired every 5~minutes during dry daylight hours. The raw disk-center radiance is measured in detector counts and averaged within $<0.1\,R_{\odot}$ of Sun center for each bandpass \citep{Penn2004SBM}. The corresponding near-Sun sky brightness is computed as the azimuthally averaged signal within the circumsolar annulus described above. This observational strategy yields a time series of disk-normalized circumsolar brightness measurements sampling a range of solar zenith angles under controlled, cloud-free conditions. A prior analysis of this dataset was conducted by \citet{tomczyk2015} in the context of evaluating daytime sky brightness at Mauna Loa for the proposed Coronal Solar Magnetism Observatory \citep[COSMO;][]{tomczyk2016}.

For this work, we adopt the SBM data products as prepared and analyzed by \citet{tomczyk2015}, rather than reprocessing the raw observations. As described by \citet{Penn2004SBM} and implemented by \citet{tomczyk2015}, instrumental scattered light is treated as a constant additive offset and removed by extrapolating the presumed linear relationship between disk-normalized sky brightness and airmass to zero airmass. The processed dataset therefore includes this correction, as well as the removal of observations affected by compromised data quality.  Here, we ascribe to each measurement an elongation that is the mean of the inner and outer FOV radii, that is, $1.54^{\circ}$ ($\sim\!6\,R_{\odot}$); however, the angular half-width is $\pm 0.77^{\circ}$ and thus each measurement represents a weighted average over a range of elongations, with the effective elongation depending on the radial variation of sky brightness.  Furthermore, each measurement is associated with the airmass along the direction to the Sun's center, thus not accounting for differences in airmass around the Sun.  Both of these limiting approximations are justified for the purposes of this study, wherein the SBM measurements serve as the reference observational constraint on circumsolar sky brightness against which aerosol-constrained inferences are evaluated in subsequent sections.

\subsection[AERONET Sun–Sky Photometry and Aerosol Inversion Products]{AERONET Sun–Sky Photometry and \\ Aerosol Inversion Products}

Mauna Loa Observatory is a primary calibration site for the Aerosol Robotic Network \citep[AERONET;][]{Holben1998}, a global network of standardized sun- and sky-scanning photometers used to characterize atmospheric aerosol properties. The standard AERONET instrument is the CIMEL Electronique CE-318 spectral radiometer, which has a full-angle field of view of $\sim\!1.2^{\circ}$. These instruments autonomously acquire direct-Sun and sky radiance measurements in discrete spectral bandpasses, including standard channels at 440, 670, 870, 935, and 1020~nm, each with a nominal bandwidth of 10~nm. While optimized for aerosol characterization rather than coronagraphy, the CE-318 collimation suppresses stray light to the $10^{-5}$ level when observing the sky at elongations of $\sim\!3^{\circ}$ from the Sun \citep{Holben1998}.

AERONET observations at Mauna Loa commenced in 1994. In this study, we limit our analysis to data acquired starting in 2000, corresponding to the adoption of a consistent operational configuration through the present day. We use both Level~1.5 direct-Sun data products and inversion products generated using the Version~3 AERONET algorithms \citep{Dubovik2000Inv, Giles2019, Sinyuk2020}. The direct-Sun measurements provide the total atmospheric optical depth (\textit{i.e.}, total extinction) as a function of wavelength, from which aerosol optical depth is retrieved after accounting for molecular (Rayleigh) scattering and gaseous absorption. The inversion products yield aerosol microphysical and optical properties, including size distributions and wavelength-dependent scattering characteristics. The specific AERONET parameters used here are summarized in Table~\ref{tbl:data_sources} and are introduced in context in subsequent sections.

The AERONET aerosol inversion products result from simultaneously fitting multi-wavelength sky radiance measurements using a plane-parallel, vertically stratified atmospheric model. These sky radiances are acquired using standardized sky-scanning geometries, most commonly the almucantar configuration, in which the polar angle is fixed at the solar zenith angle while the azimuthal angle is discretely stepped between 3 and $180^{\circ}$, to sample a wide range of scattering angles. Sky radiance scans are carried out at prescribed times based on solar zenith angle and/or elapsed time after sunrise and are completed rapidly, with each wavelength sampled on timescales of order one minute or less. Additional details regarding the inversion methodology, quality-control procedures, and data screening criteria are provided in the AERONET technical references and online project documentation.\footnote{\url{https://aeronet.gsfc.nasa.gov/new_web/Documents/Inversion_products_for_V3.png}}

In this work, we use the calibrated AERONET sky radiance products provided by the standard processing pipeline, which include instrument calibration and temperature corrections and are reported in physical units of $\mathrm{\mu W\,cm^{-2}\,sr^{-1}\,nm^{-1}}$. These sky radiance products do not include an explicit measurement of the solar disk-center radiance at the time of each scan; as a result, disk-center-normalized radiance quantities are not directly available and instead must be determined separately to facilitate comparisons to the coronagraphic observations. This is discussed further in Section~\ref{sec:model_fitting}

We again emphasize that AERONET does not directly measure sky radiance within the near-Sun circumsolar region ($\lesssim\!2^{\circ}$ from Sun center) that is most relevant for coronal observations. Instead, the retrieved aerosol optical properties provide a physically motivated basis for inferring the angular distribution of scattered solar radiation at smaller elongations through forward modeling, which we evaluate in comparison with the SBM measurements in the following sections.

\subsection{Broadband Radiation Measurements (NOAA/GML)}
\label{sec:gml}

Mauna Loa Observatory also hosts one of the NOAA Global Monitoring Laboratory (GML) Atmospheric Baseline Observatories and provides the longest continuous record of solar radiative transmission measurements \citep{Price1963, dutton2001}. In this study, we make limited use of surface downwelling measurements of direct normal irradiance (DNI) and diffuse horizontal irradiance acquired using an Eppley Normal Incidence Pyrheliometer and an Eppley Precision Spectral Pyranometer\footnote{See \url{https://gml.noaa.gov/obop/mlo/programs/esrl/solar/solar.html}.}. Both instruments measure broadband solar irradiance over the spectral range from approximately 280 to 3000~nm. These measurements provide an integrated indicator of atmospheric transparency and the partitioning of solar radiation between direct and scattered components under cloud-free conditions. Because they lack angular resolution, they are insensitive to the near-Sun circumsolar region most relevant for coronal observations; accordingly, NOAA/GML radiation data are used here only for contextual characterization of sky clarity and illumination conditions and are not employed in the quantitative inference of circumsolar sky brightness.

%%%%%%%%%%%%%%%%%%%%%%%%%%%%%%%%%%%%%%%%%%%%%%%%%%%%%%%%%%%%%%%%%%%%%%%%
%%% SECTION 3
%%%%%%%%%%%%%%%%%%%%%%%%%%%%%%%%%%%%%%%%%%%%%%%%%%%%%%%%%%%%%%%%%%%%%%%%

\begin{figure*}
    \centering
    \includegraphics[width=0.95\textwidth]{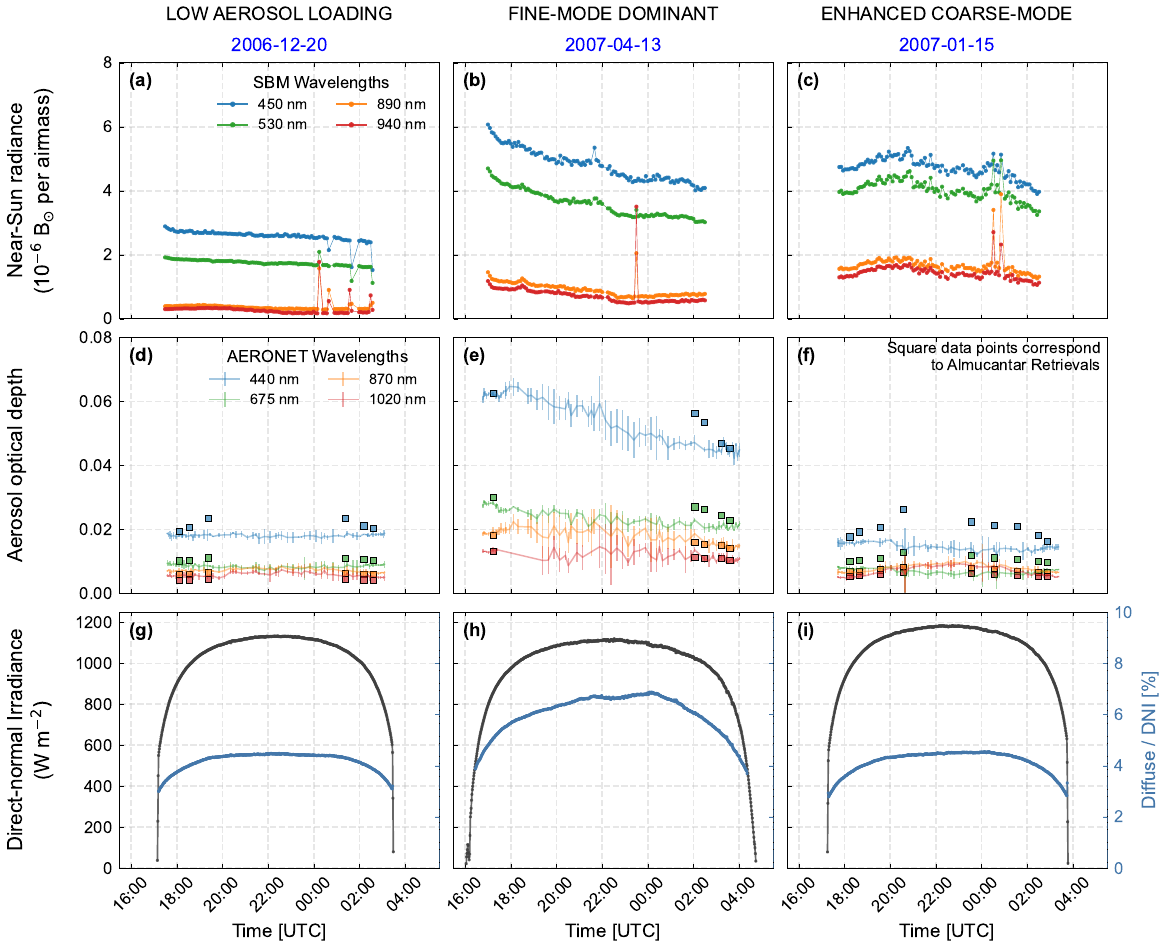}
    \caption{Diurnal evolution of circumsolar conditions on three Mauna Loa days representing Rayleigh-like (2006-12-20), fine-mode dominated (2007-04-13), and enhanced coarse-mode forward-scattering (2007-01-15) regimes. Top row: SBM near-Sun ($\Theta \sim\!1.54^{\circ}$) radiance normalized by solar disk-center radiance and corrected for airmass. Middle row: spectral AERONET aerosol optical depth; square symbols mark values inferred via the almucantar inversions. Thin error bars represent the variability of measurement triplets.  Bottom row: NOAA/GML direct-normal irradiance (black) and diffuse-to-direct ratio (blue).}
    \label{fig:compare3days}
\end{figure*}

\section{Representative Aerosol Regimes and Strengths of Circumsolar Scattering}\label{sec:daily_comparisons}

Using the observational datasets described in Section~\ref{sec:data}, we examine how circumsolar sky brightness responds to differing atmospheric conditions at Mauna Loa focusing on variations in aerosol microphysics. The aerosol regimes considered here are meant to be illustrative and they differ primarily in particle size populations. Aerosols dominated by submicron particles tend to produce weak forward scattering and angular distributions approaching the Rayleigh limit, whereas micron-scale particles strongly enhance forward scattering at small solar elongations.  Often, a bimodal distribution of `fine-mode' (sub-micron radii) and `coarse-mode' (micron-scale) particles can well describe aerosols in Earth's atmosphere \citep{oneill2001, Dubovik2000Inv}. 

\begin{figure}
    \centering
    \includegraphics[width=0.99\columnwidth]{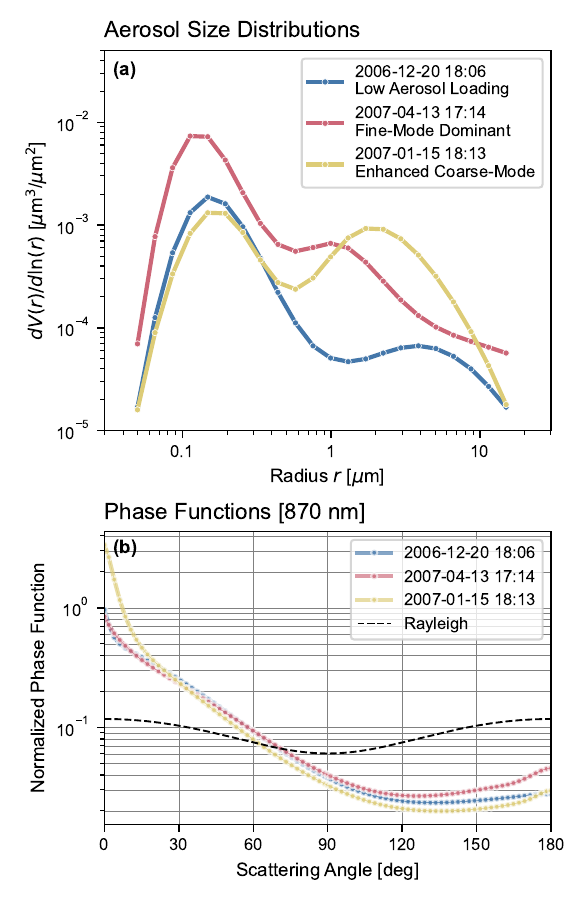}
\caption{AERONET aerosol property retrievals for the three representative cases shown in Figure~\ref{fig:compare3days}. (a) Aerosol volume size distributions at the almucantar scan times. (b) Corresponding normalized aerosol phase functions, highlighting the enhanced forward scattering in the coarse-mode case that drives the strong near-Sun radiance observed by the SBM at $\sim\!1.54^{\circ}$ ($6\,R_{\odot}$).}
    \label{fig:phase_functions}
\end{figure}

In Figures~\ref{fig:compare3days} and~\ref{fig:phase_functions}, we examine three representative days that we manually selected to illustrate distinct aerosol regimes: a Rayleigh-like case with very low aerosol loading (2006-12-20), a fine-mode–dominated case (2007-04-13), and an enhanced coarse-mode case (2007-01-15).  The mean precipitable water vapor was low on all three days ($0.49 \pm 0.04$, $2.22 \pm 0.12$, and $1.35 \pm 0.39\,\mathrm{mm}$, respectively), and all periods were cloud free. The fine-mode case coincided with elevated surface winds, suggesting enhanced mixing of boundary-layer aerosols, similar to the influence of diurnal anabatic winds that is discussed further in Section~\ref{sec:diurnal_structure}.

Figure~\ref{fig:compare3days} compares the diurnal evolution of three complementary quantities for each day. The top row shows near-Sun sky radiance measured by the SBM, averaged over an annulus between 4.5 and $7.8\,R_{\odot}$ (scattering angles $\sim\!1.12^{\circ}$–$1.95^{\circ}$). These radiances are given in units normalized by the solar disk-center radiance and further divided by the relative airmass, yielding circumsolar brightness per unit optical path. The middle row shows the spectral aerosol optical depth (AOD) from AERONET at 440, 675, 870, and 1020~nm; square markers indicate values derived during the almucantar scans used for the inversions discussed below.  We note that these are vertical optical depths, and that transmittance calculated using the Bouguer–Lambert law must include the relative air mass factor.  Finally, the bottom row shows NOAA/GML broadband direct-normal irradiance and the diffuse-to-direct ratio.

The three days exhibit distinct circumsolar behavior. The Rayleigh-like case shows weak near-Sun radiance and little temporal variability, consistent with stable molecular scattering. Measured AOD is at or below the $\sim0.01$–$0.02$ uncertainty level \citep{Dubovik2000Accuracy}. The fine-mode case exhibits elevated AOD and increased near-Sun radiance. In contrast, the coarse-mode case produces the strongest circumsolar signal observed by the SBM at infrared wavelengths despite only a weak AOD enhancement. Broadband irradiance is broadly similar across all three days, while the fine-mode case shows an enhanced diffuse-to-direct ratio indicative of wider-angle haze.

The microphysical origin of these differences can be understood from Figure~\ref{fig:phase_functions}, which shows AERONET inversion results for a representative almucantar scan. Panel~(a) presents the retrieved aerosol volume size distributions. The Rayleigh-like day shows a comparatively negligible aerosol volume, implying scattering dominated by molecular air. Aerosol volume concentration is substantially higher in the fine- and coarse-mode cases, with the fine-mode case exhibiting a pronounced accumulation peak near $\sim\!0.1$--$0.2\,\mu\mathrm{m}$ and the coarse-mode case showing a broad shoulder extending beyond $1\,\mu\mathrm{m}$.

Panel~(b) shows the corresponding $4\pi$ normalized aerosol phase functions, which corresponds to the probability density of photons scattered into a direction $\Theta$. The fine-mode case exhibits modest enhancement in side scattering relative to the Rayleigh-like case, indicating that the increased sky brightness primarily reflects increased aerosol loading as compared to strong changes in the phase function. In contrast, the coarse-mode case shows strongly enhanced forward scattering at small angles, directly accounting for the elevated near-Sun radiance measured on 2007-01-15.

In summary, Figures~\ref{fig:compare3days} and~\ref{fig:phase_functions} demonstrate that circumsolar sky brightness is controlled not solely by aerosol loading but by the angular scattering structure encoded in the aerosol phase function. AERONET almucantar inversions therefore provide essential microphysical context for interpreting near-Sun radiance measurements.

%%%%%%%%%%%%%%%%%%%%%%%%%%%%%%%%%%%%%%%%%%%%%%%%%%%%%%%%%%%%%%%%%%%%%%%%
%%% SECTION 4 
%%%%%%%%%%%%%%%%%%%%%%%%%%%%%%%%%%%%%%%%%%%%%%%%%%%%%%%%%%%%%%%%%%%%%%%%

\section{Forward Modeling of Circumsolar Radiance Using AERONET Retrievals}
\label{sec:aeronet_model}

To extend the aerosol constraints provided by AERONET into the near-Sun regime relevant for coronal observations ($\lesssim\!2^{\circ}$), we employ a constrained forward-modeling approach that uses AERONET-retrieved aerosol optical properties to predict (or infer) circumsolar radiance at the angular scales probed by the Sky Brightness Monitor (SBM). In this framework, aerosol microphysical properties are prescribed entirely by the AERONET inversion products. Additional terms serve only as effective closures for scattering contributions—such as molecular multiple scattering and ground-reflected radiation—that are already treated self-consistently within the AERONET retrievals but are not explicitly represented in the simplified analytical formulation used here.

\subsection{Analytical Almucantar Radiance Approximation}
\label{sec:almucantar_model}

We adopt a reduced analytical approximation for a uniform, plane-parallel atmospheric layer to express the almucantar (\textit{i.e.}, constant-airmass) sky radiance ($L$) at wavelength $\lambda$ and scattering angle $\Theta$ \citep[see, e.g.,][]{DeVore2012,Gueymard2001}:
\begin{equation}
\begin{split}
L_{\lambda}(\Theta) =\;
&m\, \Phi_{\lambda}\, \exp\!\left(-m\,\tau_{\mathrm{total},\lambda}\right)
\Bigg[
    \underbrace{
    \omega_{0,\lambda}\,\tau_{\mathrm{a},\lambda}\,
    \frac{P_{\mathrm{a},\lambda}(\Theta)}{4\pi}
    }_{\text{aerosol single scattering}}
\\[6pt]
&
    \,+\;
    \underbrace{
    \left(\tau_{\mathrm{R},\lambda} \,+\, \tau_{\mathrm{mR},\lambda}\right)\,
    \frac{P_{\mathrm{R}}(\Theta)}{4\pi}
    }_{\text{Rayleigh scattering with multiple-scattering correction}}
\\[6pt]
&
    \,+\;
    \underbrace{
    \tau_{\mathrm{mg},\lambda}\,
    \frac{P_{\mathrm{R}}\!\left(0^{\circ}\right)}{4\pi}
    }_{\text{ground-reflected contribution}}
\Bigg].
\end{split}
\label{eq:sky_lam}
\end{equation}
Here $m$ is the optical airmass, $\Phi_{\lambda}$ is the solar spectral irradiance at the top of the atmosphere, and $\tau_{\mathrm{total},\lambda}$ is the total atmospheric optical depth derived from AERONET direct-Sun measurements. The bracketed terms represent aerosol single scattering, molecular (Rayleigh) scattering augmented by an effective correction for multiple scattering, and a contribution from ground-reflected radiation scattered back into the line of sight. The prefactor $\exp\!\left(-m\,\tau_{\mathrm{total},\lambda}\right)$ provides an effective atmospheric transmittance based on the scan-averaged optical depth and airmass.

The solar spectrum $\Phi_{\lambda}$ for each of the AERONET bandpasses is taken as the average of the composite extraterrestrial irradiance of \citet{gueymard2004} corrected for the Earth-Sun distance at the time of observation.  All aerosol optical quantities appearing in Equation~\ref{eq:sky_lam}---including the aerosol optical depth $\tau_{a,\lambda}$, single-scattering albedo $\omega_{0,\lambda}$, and aerosol phase function $P_{\mathrm{a},\lambda}(\Theta)$---are taken directly from the AERONET inversion products and are held fixed. Molecular scattering is represented using the Rayleigh phase function \citep{Gueymard2001},
\begin{equation}
P_{\mathrm{R}}(\Theta)
= \frac{3}{4 + 2\delta}
\left(1 + \delta + (1 - \delta)\,\cos^2\!\Theta \right)
\end{equation}
where $\delta$ is the molecular depolarization factor, fixed at $\delta = 0.0279$.   The Rayleigh optical depth $\tau_{\mathrm{R},\lambda}$ is also provided in the AERONET data products as calculated based on the assumptions of \citet{Holben1998}.  The aerosol phase functions provided by AERONET are given on a fixed set of Gaussian-quadrature scattering angles (83 points; \citet{Dubovik2011}) between 0 and 180$^{\circ}$, which, for our purposes, adequately resolve the angular structure of the forward-scattered peak. Cubic spline interpolation is used here to interpolate the phase function to the angles of the measured almucantar radiances. 

\subsection{Analytical Model Fitting}
\label{sec:model_fitting}

We fit the above analytical sky-radiance model directly to the calibrated AERONET almucantar observations.  The only free parameters are the two wavelength-dependent closure terms: $\tau_{mR,\lambda}$, representing molecular multiple scattering, and $\tau_{mg,\lambda}$, representing ground-reflected radiation scattered back into the line of sight. Because the aerosol phase function is fixed by the AERONET inversion, these closure terms modify the amplitude of the Rayleigh-like background component of the modeled radiance, rather than its angular structure at small scattering angles.

For each almucantar scan and wavelength, the model parameters are obtained by minimizing a weighted least-squares merit function defined in logarithmic radiance space,
\begin{equation}
\chi^2
= \sum_i
\left[
\frac{\ln L_{\lambda,i}^{\mathrm{mod}} - \ln L_{\lambda,i}^{\mathrm{obs}}}
{\sigma_{\ln L}}
\right]^2
\end{equation}
where $L_{\lambda,i}^{\mathrm{obs}}$ and $L_{\lambda,i}^{\mathrm{mod}}$ are the observed and modeled sky radiances at scattering angle $i$, and $\sigma_{\ln L}$ represents the fractional uncertainty in the measured radiances. We adopt $\sigma_{\ln L} = 0.05$ motivated by the typical residual levels reported for AERONET Version~3 almucantar inversions, which reflects the combined effects of measurement noise, calibration uncertainty, and model approximation \citep[e.g.,][]{Dubovik2000Accuracy,Giles2019}. Minimization is performed using a differential-evolution algorithm, with bounds chosen to ensure physically plausible parameter values.  A more detailed discussion of the distribution of the retrieved closure-term values is provided in Appendix \ref{sec:appendix1}. With regards to data quality assurance and outlier rejection, we follow the fitting philosophy employed in the AERONET inversion framework. In particular, only scattering angles larger than $3^{\circ}$ are included and symmetric almucantar samples must agree within 20\%.

Once again following the AERONET methodology, we calculate the goodness of fit value for the model as an \emph{optical residual error}, defined as
\begin{equation}
\epsilon_{\mathrm{opt}}
= \sigma_{\ln L}
\sqrt{\frac{\chi^2}{N_{\mathrm{eff}} - N_p}}
\end{equation}
where $N_{\mathrm{eff}}$ is the effective number of retained angular samples and $N_p$ is the number of fitted parameters. Fits with $\epsilon_{\mathrm{opt}} \lesssim 5\%$ are retained for further analysis, consistent with the residual thresholds typically achieved in AERONET almucantar inversions.

Once determined for each individual scan, $\tau_{\mathrm{mR},\lambda}$ and $\tau_{\mathrm{mg},\lambda}$ are held fixed, and then the model radiance is evaluated at a scattering angle of $1.54^{\circ}$ for comparison with the independent SBM measurements. The calculated radiances are converted to units normalized to the extinction-attenuated mean solar radiance where irradiance is converted to radiance using an ephemeris value for the apparent solar diameter.  To ensure consistency with the SBM measurements, which are normalized to the solar disk-center radiance, we apply a wavelength-dependent correction (accounting for solar limb darkening) that converts mean-disk to disk-center radiance using tabulated values from \citet{cox2000}. This correction is applied uniformly to all modeled radiances expressed relative to the solar radiance.

\subsection{Comparison with Almucantar and Near-Sun Observations}
\label{sec:model_results}

\begin{figure*}
    \centering
    \includegraphics[width=0.95\textwidth]{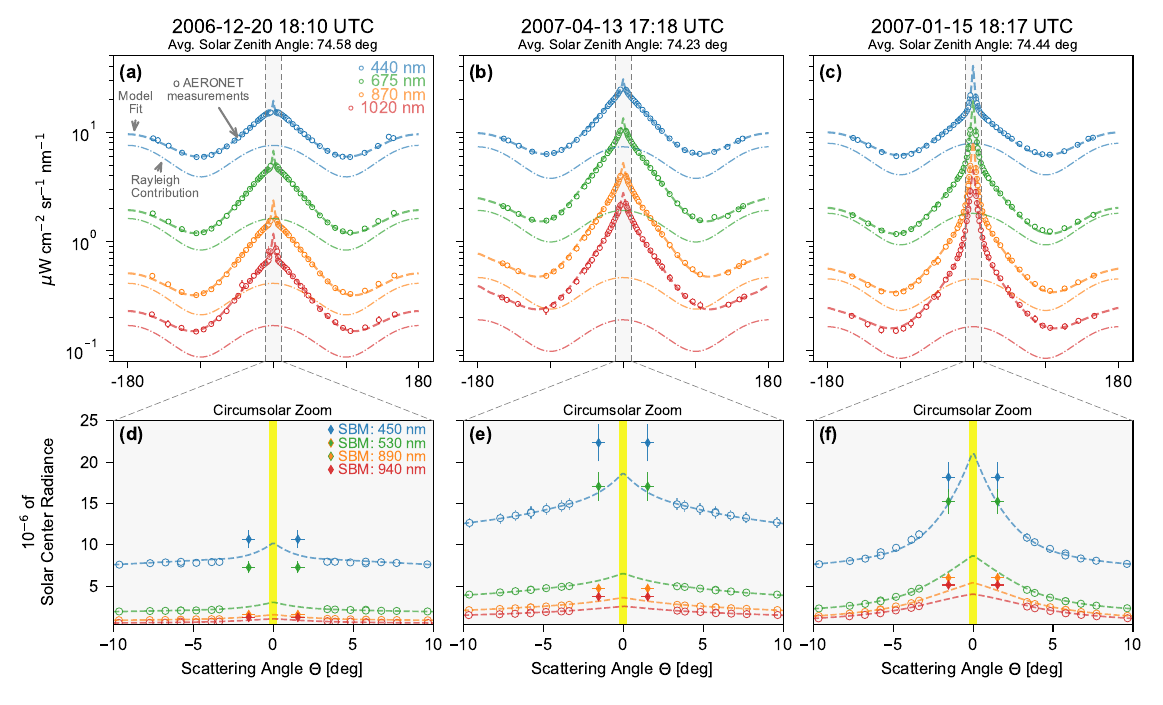}
\caption{Comparison of measured AERONET almucantar sky radiances (circles) and analytical model radiances from Equation~\ref{eq:sky_lam} (dashed curves) for three representative aerosol regimes at Mauna Loa: a Rayleigh-like day (2006-12-20), a fine-mode--dominated day (2007-04-13), and a coarse-mode forward-scattering day (2007-01-15). Panels (a)–(c) show radiances over the full range of scattering angles at four wavelengths, with the Rayleigh contribution indicated separately (dash-dotted lines). Panels (d)–(f) show an expanded view of the near-Sun region with a linear scaled y-axis in relative radiance units. Here, the independent co-temporal Sky Brightness Monitor (SBM) measurements are overplotted (diamonds).}
    \label{fig:almu_rad}
\end{figure*}

For the three representative aerosol regimes introduced in Section~\ref{sec:daily_comparisons}, Figure~\ref{fig:almu_rad} compares sky radiances modeled using Equation~\ref{eq:sky_lam} with measured AERONET almucantar radiances over the full range of scattering angles sampled by each scan. In all cases, the model reproduces both the magnitude and angular structure of the observed radiances across wavelengths, with Rayleigh scattering dominating at large angles and aerosol scattering becoming increasingly important toward smaller angles.

Panels~(d)–(f) of Figure~\ref{fig:almu_rad} focus on the near-Sun region ($|\Theta| \lesssim\!10^{\circ}$), where the fitted model radiances can be compared directly with SBM measurements. In this regime, the angular dependence of sky radiance is highly nonlinear, particularly for aerosol populations with strong forward scattering, such that simple empirical extrapolation of almucantar radiances toward small scattering angles is inadequate. The analytical model in Equation~\ref{eq:sky_lam} provides a physically constrained extension of the AERONET retrievals into this regime by enforcing consistency with the aerosol phase function. The resulting agreement demonstrates that the model successfully bridges the angular gap between the AERONET and SBM measurements without altering the aerosol microphysical inputs.  We comment further on the remaining discrepancies of the two approaches below. 

\begin{figure*}
    \centering
    \includegraphics[width=0.95\textwidth]{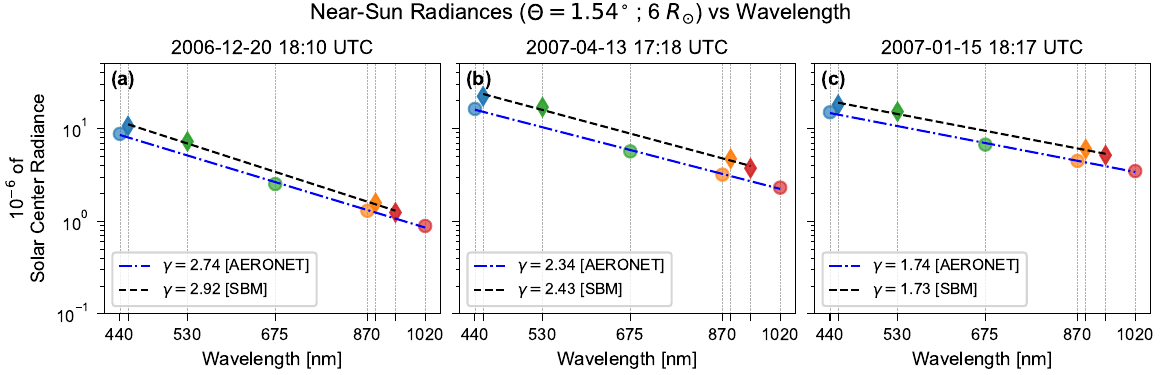}
\caption{Wavelength dependence of near-Sun radiances at $\Theta \approx 1.54^{\circ}$ ($\approx 6\,R_{\odot}$) for the same three representative days shown in Figure~\ref{fig:almu_rad}. AERONET-inferred radiances (circles) and SBM measurements (triangles) are normalized by the solar disk-center radiance and corrected for airmass. Dashed lines indicate best-fit power-law slopes, parameterized by $\gamma$.}
    \label{fig:gamma}
\end{figure*}

Figure~\ref{fig:gamma} summarizes the wavelength dependence of the circumsolar radiance at $\Theta \approx 1.54^{\circ}$ ($\approx 6\,R_{\odot}$). For each representative day, near-Sun radiances inferred from the AERONET-based model and measured by the SBM are normalized by the solar disk-center radiance, corrected for airmass, and parameterized as a power law, $L_\lambda \propto \lambda^{-\gamma}$. The parameter $\gamma$ provides a compact diagnostic of the chromatic behavior of circumsolar scattering. Despite differences in the absolute radiance amplitudes, which we attribute to variances due to instrument FOV, angular averaging, and modeling approximations (discussed further below), the inferred spectral slopes from the AERONET-based model and the SBM measurements are in close agreement, indicating that the constrained forward model captures the dominant wavelength dependence of near-Sun scattering.

Together, the agreement in angular structure and spectral slope establishes a physically meaningful connection between AERONET aerosol retrievals and near-Sun sky brightness measurements. Having demonstrated this consistency, we next examine how the inferred circumsolar radiance and its spectral dependence vary across the full observational dataset and relate to aerosol optical properties.

%%%%%%%%%%%%%%%%%%%%%%%%%%%%%%%%%%%%%%%%%%%%%%%%%%%%%%%%%%%%%%%%%%%%%%%%
%%% SECTION 5
%%%%%%%%%%%%%%%%%%%%%%%%%%%%%%%%%%%%%%%%%%%%%%%%%%%%%%%%%%%%%%%%%%%%%%%%

\section{Statistical Comparison of AERONET-Inferred and SBM-Measured Circumsolar Brightness}\label{sec:statistics}

Here we examine the statistical behavior of the circumsolar sky brightness using the combined AERONET–SBM dataset. We assess both the inter-approach consistency and the inner-parameter relationships. 

\subsection{Radiometric Consistency Between Instruments}
\label{sec:rad_consistency}

A first-order test of the forward-modeling framework is whether circumsolar radiances inferred from AERONET aerosol retrievals reproduce the independently measured near-Sun sky brightness from the Sky Brightness Monitor (SBM). Although the two systems differ fundamentally in measurement strategy, angular sampling, and calibration approach, both probe the same underlying forward-scattering aerosol physics. Here we assess their radiometric consistency at a common scattering angle of $\Theta \approx\!1.54^{\circ}$.

\begin{figure}
    \centering
    \includegraphics[width=0.95\columnwidth]{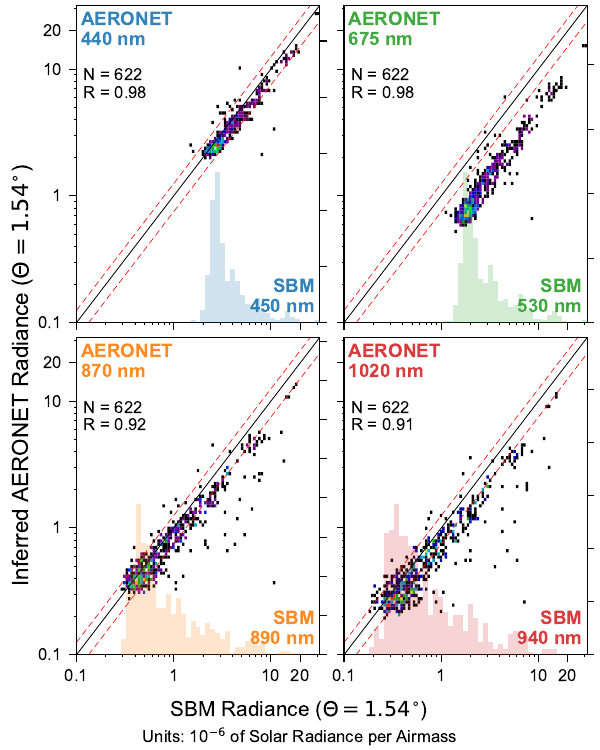}
    \caption{Two-dimensional histograms comparing AERONET-inferred circumsolar radiance at $\Theta \approx 1.54^{\circ}$, derived from sky-radiance inversions of almucantar measurements, with contemporaneous SBM near-Sun radiance measurements for four wavelength pairs. Radiances are expressed in units of $10^{-6}$ of the solar disk-center radiance per unit airmass. Each panel corresponds to one AERONET channel paired with the nearest SBM wavelength. Logarithmic binning emphasizes the dynamic range of the measurements. The solid line indicates the one-to-one relation whereas the dashed red lines show a $\pm$ 25\% deviations from unity.  One-dimensional histograms of the SBM measurements are shown in the background.}
    \label{fig:aero_vs_sbm}
\end{figure}

Figure~\ref{fig:aero_vs_sbm} compares AERONET-inferred near-Sun radiances with contemporaneous SBM measurements on a point-by-point basis over the full year of overlapping observations. Two-dimensional histograms are shown for each wavelength, pairing each AERONET channel with the nearest SBM band. Across all wavelengths, the correspondence is strong, with rank correlation coefficients exceeding 0.9, indicating that the AERONET-based model captures the day-to-day variability in circumsolar brightness measured by the SBM.

For wavelength pairs that are closely matched, the distributions cluster near the one-to-one relation, while larger systematic offsets appear when the paired wavelengths differ by several tens of nanometers. At longer wavelengths, increased statistical scatter reflects both reduced signal levels and greater sensitivity to uncertainties in the forward-scattering portion of the aerosol phase function.  As consistent with Figure~\ref{fig:gamma}, the SBM values are systematically $\sim 5$ to $10\%$ larger than those inferred with AERONET.  Relative to the requirements of coronal sky-quality assessment, these differences are of limited practical impact; however, we also anticipate that additional gains in consistency may be achieved through more detailed quantification of the remaining model and measurement uncertainties.  From a measurement standpoint, the small offsets from the unity relation and the observed scatter likely reflect a combination of factors, including differences in effective wavelength, limits on the absolute radiance accuracy and sensitivity, and the quality of the SBM’s empirical correction for instrumental stray light. Also, as mentioned previously, we do not attempt a more complex correction for the $\pm 0.77^{\circ}$ radial angular half-width of the SBM FOV, electing instead to compare at a single elongation value.  This simplification can bias the relationships when the aerosol phase function is strongly forward-peaked. However, Figure~\ref{fig:almu_rad} (panel f) indicates that, for the aerosol loading conditions observed at Mauna Loa, the near-Sun radiance gradient remains modest relative to the SBM measurement uncertainties. Overall, despite these limitations, the favorable agreement demonstrates that the AERONET-constrained forward model reproduces circumsolar radiance variability with reasonable fidelity across the dynamic range of the Mauna Loa conditions. 

\subsection{Information Content of Near-Sun Radiance Gradient}
\label{sec:angular_stats}

\begin{figure}
    \centering
    \includegraphics[width=0.95\columnwidth]{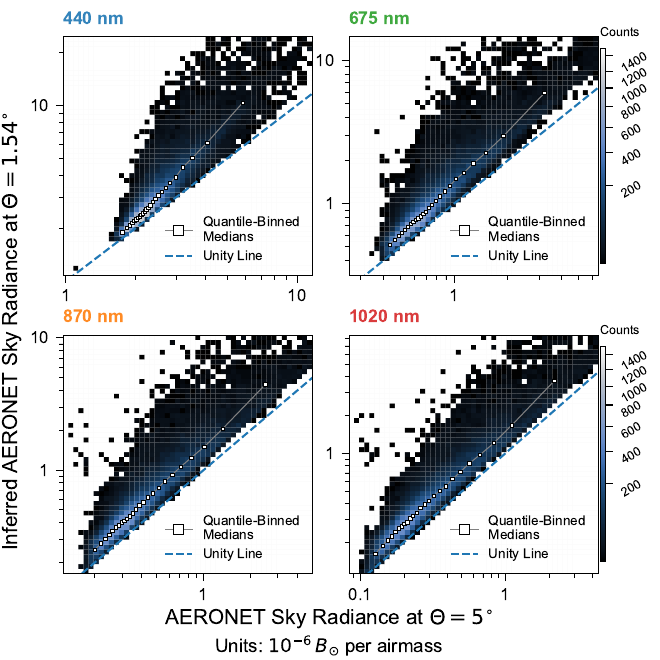}
    \caption{Two-dimensional histograms of AERONET-inferred sky radiance at $\Theta \approx 1.5^{\circ}$ versus $\Theta = 5^{\circ}$ for four wavelengths. Colors indicate measurement density in log--log space; the dashed line marks equality between scattering angles, and white curves show quantile-binned medians.}
    \label{fig:angular_comparison}
\end{figure}

Having demonstrated the consistency above, we can further examine how the AERONET sky radiance measurements at larger scattering angles relate to the near-Sun region that is not directly measured by AERONET.  Figure~\ref{fig:angular_comparison} compares AERONET-inferred sky radiances at $\Theta \approx 1.5^{\circ}$ with the measurements at $\Theta = 5^{\circ}$ across all wavelengths and observations.  Two-dimensional histograms and quantile-binned medians are shown in log--log space. If radiance scaled weakly with scattering angle, measurements at $5^{\circ}$ would provide a reliable proxy for near-Sun brightness. Instead, substantial departures from the one-to-one relation are evident at all wavelengths.

Inferred radiances at $\Theta \approx 1.5^{\circ}$ consistently exceed those at $5^{\circ}$, with enhancements spanning more than an order of magnitude. This behavior reflects the steep rise of the aerosol phase function toward small scattering angles and demonstrates that sky radiance measured several degrees from the Sun does not uniquely predict circumsolar brightness within $\sim$2$^{\circ}$. The largest deviations should occur under coarse-mode--dominated conditions given that fine-mode or Rayleigh-dominated cases exhibit more modest angular enhancement.

The relationship is wavelength dependent. At shorter wavelengths (440 and 675~nm), the spread between $\Theta \approx 1.5^{\circ}$ and $5^{\circ}$ radiances is largest, whereas at longer wavelengths (870 and 1020~nm) the correspondence tightens, though substantial scatter remains for coarse-mode cases. These results show that extrapolation into the inner circumsolar regime captures information that is not available from larger-angle measurements alone and motivates the use of aerosol-constrained forward modeling to interpret near-Sun sky brightness.

\subsection{Common Aerosol Proxies for Circumsolar Sky Quality}
\label{sec:proxy_params}

\begin{figure}
    \centering
    \includegraphics[width=0.99\columnwidth]{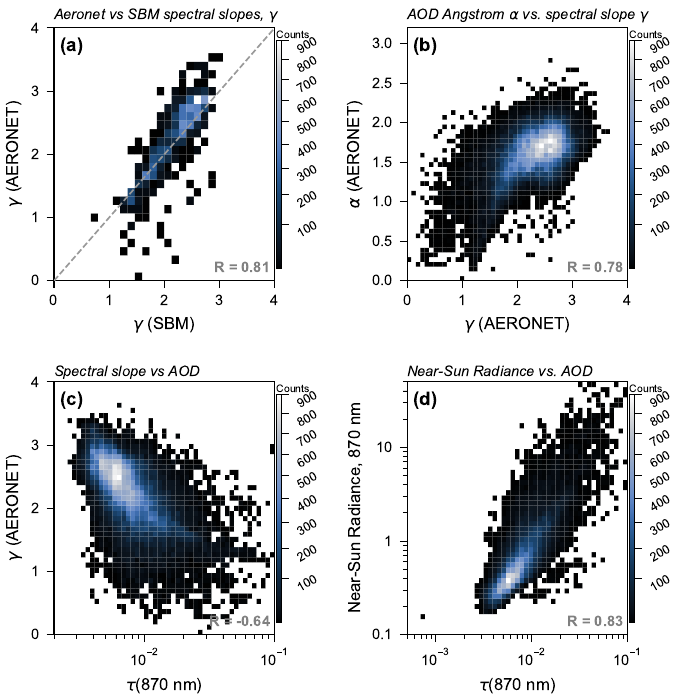}
\caption{Relationships between near-Sun radiance, spectral slope, and common aerosol proxy parameters derived from AERONET and SBM observations.
(\textit{a}) Comparison of the near-Sun spectral slope $\gamma$ inferred from AERONET-based modeling and from SBM measurements.
(\textit{b}) AERONET Ångström exponent $\alpha$ versus the AERONET near-Sun spectral slope $\gamma$.
(\textit{c}) Dependence of $\gamma$ on aerosol optical depth at 870~nm.
(\textit{d}) Near-Sun radiance at 870~nm as a function of aerosol optical depth at the same wavelength.
Panels~(b)–(d) include all AERONET almucantar model fits from 2000 to 2025. Colors indicate the relative density of measurements. Pearson correlation coefficients are given in lower right of each panel.}
    \label{fig:gammaStats}
\end{figure}

Building on the wavelength-dependence analysis in Figure~\ref{fig:gamma}, Figure~\ref{fig:gammaStats}a compares the near-Sun spectral slope $\gamma$ inferred independently from SBM observations and from AERONET-inferred radiances at $\Theta \approx 1.54^{\circ}$. The two estimates agree closely (Pearson correlation coefficient, $R$, is $0.81$), indicating that the wavelength dependence of circumsolar scattering is robustly captured by the AERONET-constrained inferences.

We next extend the analysis of the AERONET-based inferences (2000--2025) to examine commonly used column-integrated aerosol proxies for daytime sky conditions, including aerosol optical depth (AOD) and the Ångström exponent $\alpha$. The Ångström exponent \citep{angstrom1929} parametrizes the approximate power-law wavelength dependence of aerosol extinction and is widely used as a qualitative indicator of particle size: larger values generally correspond to fine-mode--dominated aerosols, whereas smaller values indicate a stronger coarse-mode contribution. We emphasize that this interpretation remains qualitative, as the mapping from $\alpha$ to microphysical properties is not unique \citep{schuster2006}. Figures~\ref{fig:gammaStats}b--d then relate $\alpha$ to $\gamma$ (panel~b), $\gamma$ to AOD (panel~c), and the inferred near-Sun radiance to AOD (panel~d).

Systematic tendencies are evident. In Figure~\ref{fig:gammaStats}b, the AERONET Ångström exponent $\alpha$ increases with the AERONET-derived near-Sun spectral slope $\gamma$ ($R=0.79$), consistent with both quantities responding to column-integrated particle size and the resulting chromaticity of circumsolar scattering. In contrast, $\gamma$ decreases with increasing AOD at 870~nm (Figure~\ref{fig:gammaStats}c; $R=-0.64$), while the inferred near-Sun radiance at 870~nm increases strongly with AOD (Figure~\ref{fig:gammaStats}d; $R=0.83$). Notably, substantial dispersion is present in these relationships: similar AOD or $\alpha$ values can correspond to markedly different near-Sun radiance levels, reflecting sensitivity to the aerosol size distribution and forward-scattering phase function that is not captured by any single bulk parameter.

Overall, these comparisons show that AOD, the Ångström exponent, and the near-Sun spectral slope $\gamma$ provide useful context for circumsolar sky quality but remain incomplete predictors of near-Sun radiance. In the following subsection, we therefore focus directly on aerosol size distributions and volume concentrations, which provide a more physically grounded interpretation of the observed variability in circumsolar radiance.

\subsection{Dependence of Near-Sun Radiance on Aerosol Coarse-Mode Volume Concentration}
\label{sec:size_dependence}

\begin{figure*}
    \centering
    \includegraphics[height=0.5\textwidth]{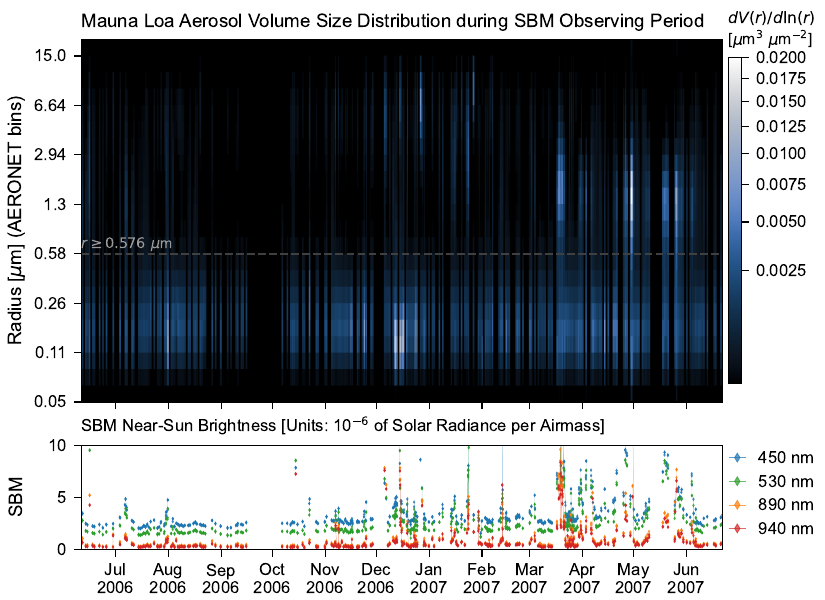}
    \includegraphics[height=0.5\textwidth]{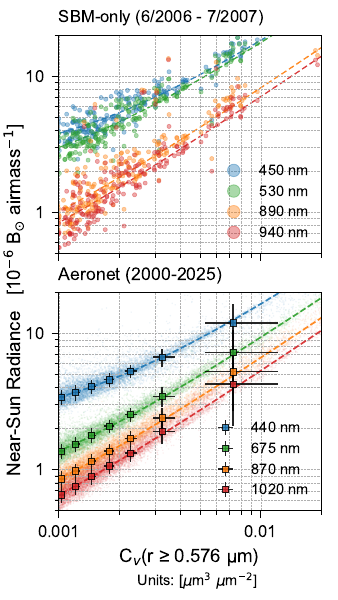}
\caption{
(\textit{Left}) AERONET aerosol volume size distributions at Mauna Loa during the SBM observing period (June~2006--July~2007), shown as $\mathrm{d}V(r)/\mathrm{d}\ln r$, with contemporaneous SBM near-Sun sky brightness measurements (lower panel) normalized by airmass.
(\textit{Right}) Near-Sun radiance at $\Theta \approx\!1.54^{\circ}$ versus coarse-mode aerosol volume concentration, $C_v(r \ge 0.576\,\mu\mathrm{m})$. The upper panel shows SBM campaign measurements with robust linear fits, while the lower panel shows the Mauna Loa AERONET data (2000--2025) with binned medians and weighted linear fits.}
    \label{fig:size_dist}
\end{figure*}

\begin{deluxetable}{lcccc}
\tabletypesize{\footnotesize}
\tablecaption{Linear fits of near-Sun radiance versus coarse-mode volume concentration\label{tab:cv_fits}}
\tablehead{
\colhead{Dataset} & \colhead{$\lambda$ (nm)} & \colhead{Slope $m$} & \colhead{Intercept $b$} & \colhead{Method}}
\startdata
SBM & 450 & 1665.172 & 2.152 & RANSAC \\
SBM & 530 & 1618.456 & 1.384 & RANSAC \\
SBM & 890 & 818.835 & 0.213 & RANSAC \\
SBM & 940 & 698.547 & 0.152 & RANSAC \\
AERONET & 440 & 1462.128 & 1.938 & WLS \\
AERONET & 675 & 893.807 & 0.474 & WLS \\
AERONET & 870 & 645.465 & 0.209 & WLS \\
AERONET & 1020 & 522.040 & 0.129 & WLS \\
\enddata
\tablecomments{Fits correspond to $I_{\mathrm{sky}} = m\,C_v(r \geq 0.576~\mu\mathrm{m}) + b$, where $I_{\mathrm{sky}}$ is the near-Sun radiance at $\Theta \approx 1.54^{\circ}$ expressed in units of $10^{-6}\,B_{\odot}$ per airmass, and $C_v$ is the coarse-mode volume concentration in $\mu\mathrm{m}^{3}\,\mu\mathrm{m}^{-2}$ (equivalently, $\mu$m). The slope $m$ therefore has units of $(10^{-6}\,B_{\odot}\,\mathrm{airmass}^{-1})/(\mu\mathrm{m}^{3}\,\mu\mathrm{m}^{-2})$, and $b$ has units of $10^{-6}\,B_{\odot}$ per airmass.}
\end{deluxetable}

The proxy relationships in Section~\ref{sec:proxy_params} indicate that bulk extinction metrics do not uniquely predict circumsolar sky brightness. This is expected because near-Sun radiance depends most strongly on the aerosol phase function at small scattering angles, which is controlled by the detailed particle size distribution and, secondarily, by composition and particle shape.

Figure~\ref{fig:size_dist} (left) provides a heuristic illustration of this link during the SBM observing period (June~2006--July~2007). The AERONET-retrieved volume size distributions show substantial variability in both the amplitude and shape of the coarse mode. Periods with enhanced coarse-mode volume (i.e., increased $dV/d\ln r$ at radii $\gtrsim 1~\mu$m) tend to coincide with elevated SBM near-Sun radiance (lower panel, normalized by airmass), consistent with the disproportionate contribution of large particles to forward scattering at $\Theta \approx 1.54^{\circ}$. Conversely, intervals dominated by fine-mode variability often show comparatively muted changes in the near-Sun radiance despite changes in overall aerosol loading.

Motivated by this distribution-level behavior, we quantify the relationship using a size-resolved scalar measure derived directly from the AERONET volume size distribution, $\mathrm{d}V(r)/\mathrm{d}\ln r$. The column-integrated aerosol volume concentration is defined as
\begin{equation}
C_v = \int_{r_{\min}}^{r_{\max}} \frac{\mathrm{d}V(r)}{\mathrm{d}\ln r}\,\mathrm{d}\ln r
\end{equation}
where $r_{\min}$ and $r_{\max}$ span the AERONET inversion bins. Because the inner circumsolar radiance is dominated by the most forward-scattering particles, we focus on the coarse-mode volume concentration, $C_v(r \ge 0.576\,\mu\mathrm{m})$, obtained by restricting the integral to radii larger than the nominal fine--coarse separation adopted in the AERONET inversion.

Figure~\ref{fig:size_dist} (right) relates the near-Sun radiance at $\Theta \approx\!1.54^{\circ}$ to $C_v(r \ge 0.576\,\mu\mathrm{m})$. For the SBM campaign (upper panel), the radiance increases approximately linearly with coarse-mode volume at all four SBM wavelengths. We summarize robust linear fits, derived with a RANdom SAmple Consensus (RANSAC) method \citep{FischlerBolles1981RANSAC}, in Table~\ref{tab:cv_fits}; these capture the dominant scaling while limiting sensitivity to sparse, high-leverage conditions associated with unusually strong coarse-mode events.

The same first-order scaling persists when extended to the full Mauna Loa AERONET record (2000--2025; lower panel). The binned medians follow a near-linear trend over a broader dynamic range than the SBM campaign alone, and weighted least-squares fits (Table~\ref{tab:cv_fits}) provide a compact empirical mapping between coarse-mode volume and near-Sun radiance at the AERONET wavelengths. Across both datasets, the fitted slope decreases toward longer wavelengths, consistent with reduced scattering efficiency in the near-IR for a fixed particle population.

Some dispersion remains about the mean relation between the near-Sun radiance and the coarse-mode aerosol volume concentration.  Importantly, this spread is physically expected: for a fixed integrated coarse-mode volume, near-Sun radiance can vary with the detailed shape of the coarse-mode distribution (e.g., effective radius, width, and the relative contribution of $r\sim1$--$3~\mu$m particles), as well as with particle nonsphericity and refractive-index variability, all of which modulate the forward-scattering phase function. 

Overall, Figure~\ref{fig:size_dist} and Table~\ref{tab:cv_fits} demonstrate that coarse-mode volume concentration provides a more physically grounded predictor of near-Sun radiance than AOD or the Ångström exponent alone.

\section{Temporal Variability of Circumsolar Brightness}
\label{sec:temporal_variability}

Having established the aerosol properties that govern circumsolar brightness, we now examine how near-Sun sky conditions at Mauna Loa vary on interannual, seasonal, and diurnal timescales using the AERONET-inferred near-Sun radiances. 

\subsection{Multiyear and Seasonal Variability}

\begin{figure*}
    \centering
    \includegraphics[width=0.95\textwidth]{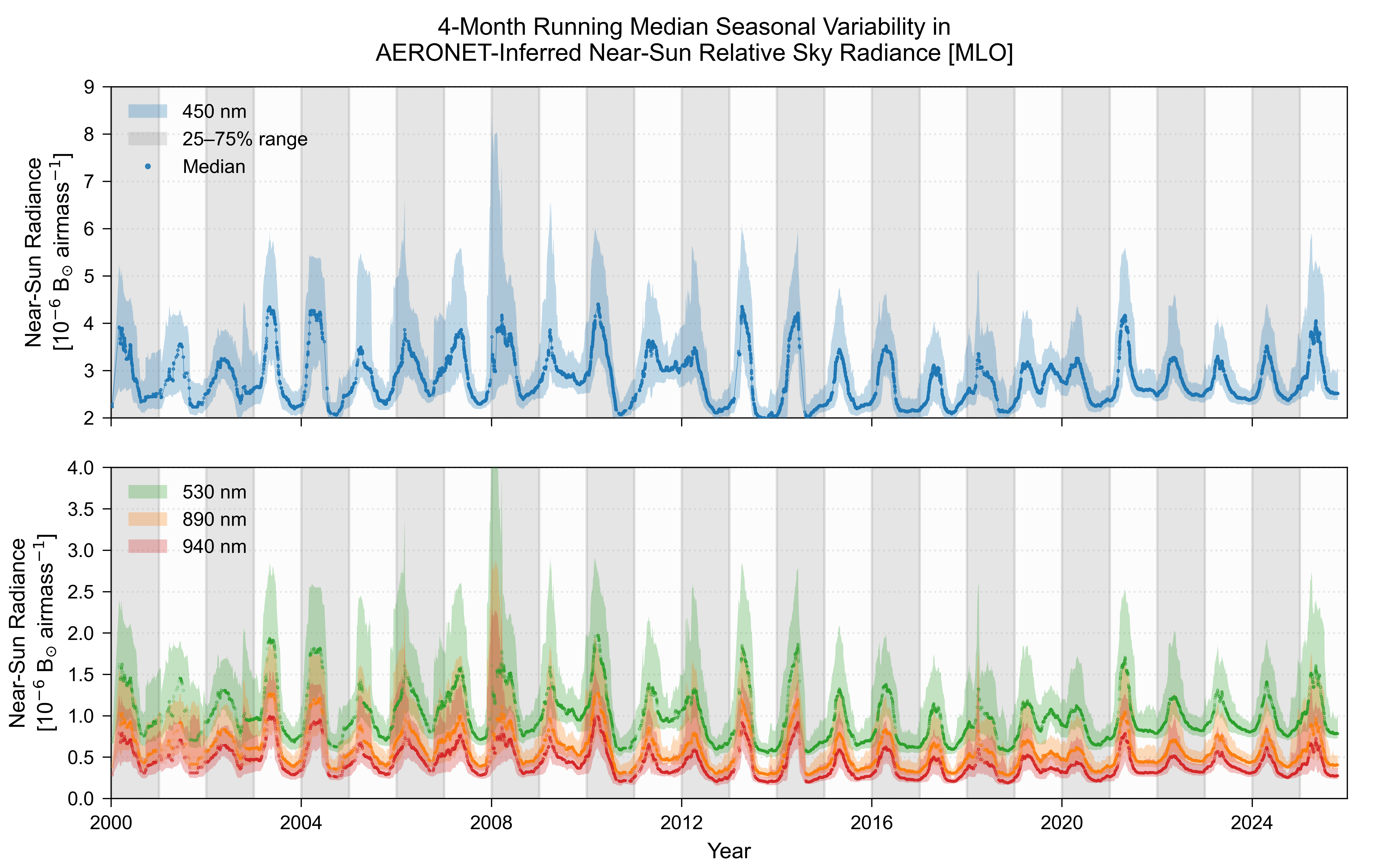}
    \caption{Multi-year variability of AERONET-inferred near-Sun sky radiance at Mauna Loa Observatory, expressed as radiance divided by airmass. Solid curves show 4-month-wide running medians, and shaded regions indicate the interquartile range. Alternating gray bands mark individual years. Shorter wavelengths exhibit larger seasonal amplitudes, reflecting increased sensitivity to molecular and fine-mode aerosol scattering.}
    \label{fig:multiyear}
\end{figure*}

\begin{figure}
    \centering
    \includegraphics[width=0.95\columnwidth]{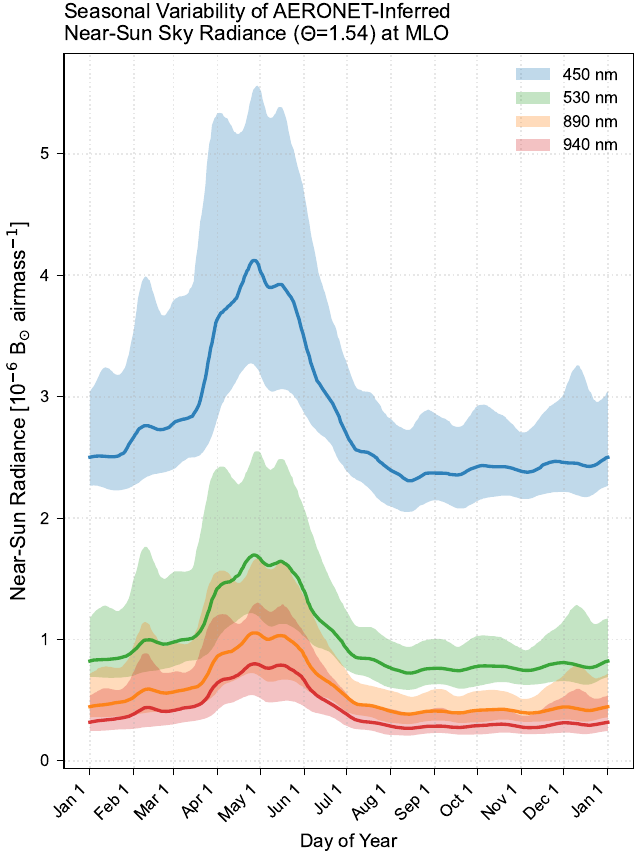}
    \caption{Seasonal climatology of AERONET-inferred near-Sun sky radiance at Mauna Loa, formed by mapping the multi-year dataset into day-of-year space and applying a Gaussian-weighted (FWHM of 14 days) running median. Shaded regions denote the interquartile range. The pronounced late-spring maximum reflects seasonal enhancement of aerosol scattering associated with long-range aerosol transport.}
    \label{fig:seasonal}
\end{figure}

Seasonal modulation is known to be a dominant feature of the circumsolar brightness record at Mauna Loa and is evident in both the multi-year time series (Figure~\ref{fig:multiyear}) and the day-of-year climatology (Figure~\ref{fig:seasonal}). The inferred near-Sun radiance per airmass exhibits a robust late-spring maximum, most pronounced at shorter wavelengths, and a broad minimum during boreal autumn and early winter.

Gaussian-weighted running medians show that circumsolar radiance peaks consistently between April and June across wavelengths, with occasional secondary enhancements associated with episodic aerosol events. When the data are mapped into day-of-year space, this behavior sharpens into a coherent climatological pattern: radiance increases rapidly in spring, declines through summer, and remains comparatively low during winter.

This seasonal cycle closely follows the established aerosol climatology of Mauna Loa. Numerous studies have linked springtime reductions in atmospheric transmission and enhanced scattering to long-range transport of Asian dust and pollution aerosols superimposed on otherwise clean free-tropospheric conditions \citep[e.g.,][]{dutton2001,perry1999,park2020}. Column-integrated AERONET observations further show that springtime increases in aerosol optical depth and shifts in fine–coarse mode partitioning accompany trans-Pacific export of both mineral dust and anthropogenic aerosols \citep[e.g.,][]{eck2005}. The seasonal maxima in near-Sun radiance inferred here occur during precisely this April–June period of enhanced Asian outflow, while wintertime minima coincide with reduced long-range transport and weaker aerosol scattering.  Year-on-year variations are influenced by weather patterns in the Pacific, including the El Niño and La Niña phases of the El Niño-Southern Oscillation (ENSO).

\subsection{Diurnal Structure}\label{sec:diurnal_structure}

\begin{figure}
    \centering
    \includegraphics[width=0.95\columnwidth]{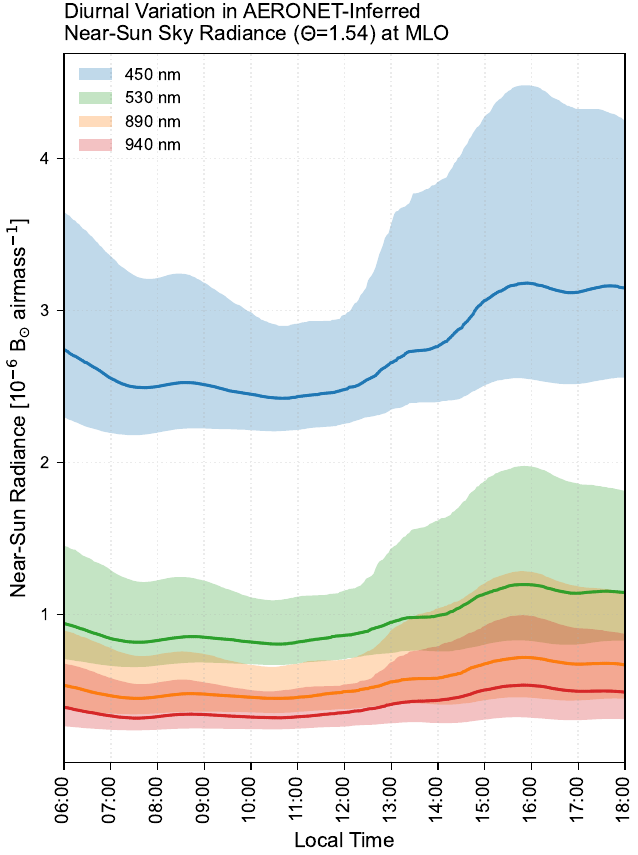}
    \caption{Diurnal variation of AERONET-inferred near-Sun sky radiance at Mauna Loa, obtained by collapsing all years of data into local-time space and applying a Gaussian-weighted (FWHM of 1.2 hours) running median. Solid curves show median behavior and shaded regions indicate the interquartile range, highlighting a systematic afternoon enhancement across wavelengths.}
    \label{fig:diurnal}
\end{figure}

In addition to seasonal variability, circumsolar brightness at Mauna Loa exhibits a recurring diurnal pattern. To characterize this behavior, the full AERONET-inferred radiance record is collapsed into local-time space and smoothed using Gaussian-weighted running medians (Figure~\ref{fig:diurnal}).

Across wavelengths, near-Sun radiance increases systematically during afternoon hours. This behavior likely reflects a combination of up-slope flow and boundary-layer growth introducing additional local-sourced aerosols into the line of sight \citep{park2020}. The diurnal modulation is strongest at shorter wavelengths (450–530~nm), consistent with enhanced sensitivity to Rayleigh scattering and fine-mode aerosols, while longer wavelengths show weaker but still coherent trends.

Despite the clear median structure, the interquartile spread remains large at all wavelengths, underscoring the substantial variability inherent to a free-tropospheric site intermittently influenced by boundary-layer air and long-range transport. These results indicate that diurnal phase must be considered when interpreting circumsolar radiance measurements or using them as proxies for observing conditions, particularly for afternoon observations.

%%%%%%%%%%%%%%%%%%%%%%%%%%%%%%%%%%%%%%%%%%%%%%%%%%%%%%%%%%%%%%%%%%%%%%%%
%%% SECTION 6
%%%%%%%%%%%%%%%%%%%%%%%%%%%%%%%%%%%%%%%%%%%%%%%%%%%%%%%%%%%%%%%%%%%%%%%%

\begin{figure*}
    \centering
    %\begin{interactive}{animation}{synthetic_sky_sunrise.mp4}
\includegraphics[width=0.95\textwidth]{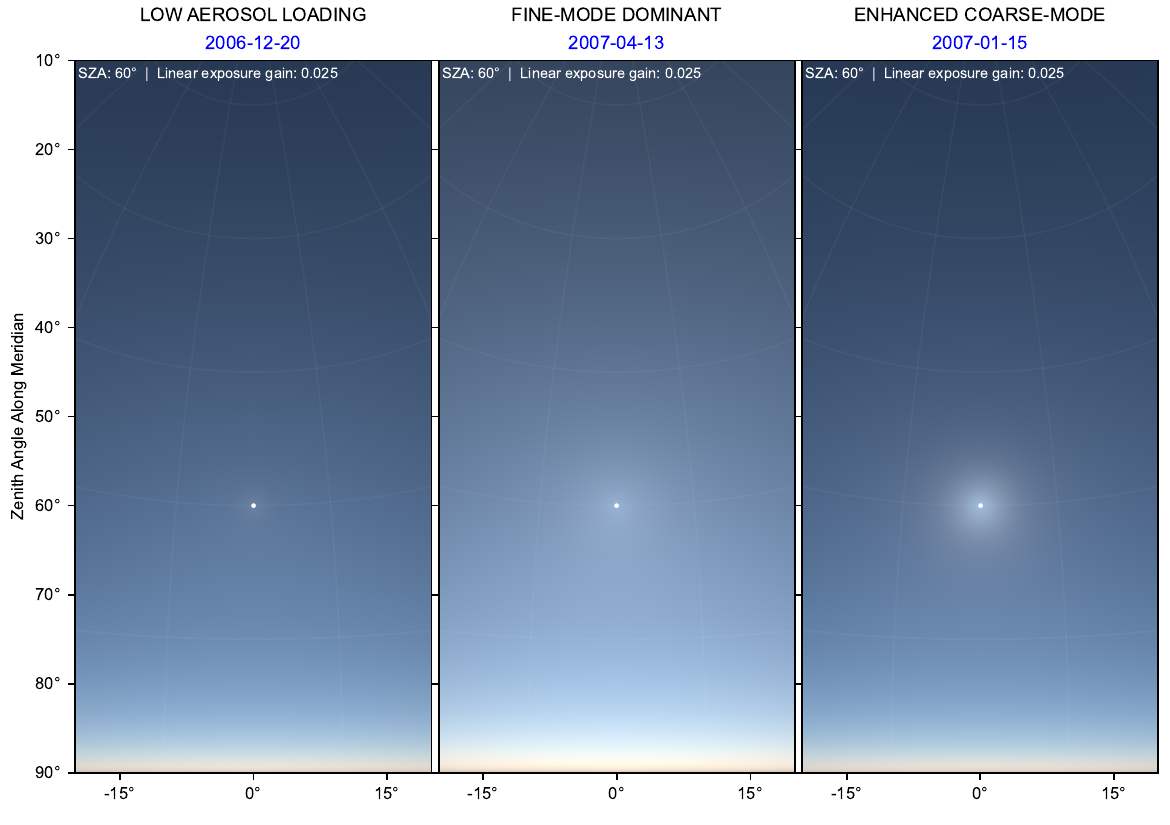}
\caption{Synthetic true-color circumsolar sky images generated using wavelength-resolved radiative-transfer modeling (libRadtran/DISORT) constrained by AERONET aerosol retrievals for three representative Mauna Loa aerosol regimes: Rayleigh-like, fine-mode–dominated, and coarse-mode forward-scattering.  Faint grid lines corresponding to contours of constant altitude and azimuth angles on sky have been added. The images illustrate how differences in aerosol microphysics manifest in the angular structure and apparent brightness of the circumsolar aureole. An animated version of this figure, showing the evolution of the synthetic sky with changing solar elevation, is available in the online journal.}
    %\end{interactive}
    \label{fig:synth_sky}
\end{figure*}

\begin{figure*}
    \centering
    %\begin{interactive}{animation}{synthetic_xchroma_sunrise.mp4}
    \includegraphics[width=0.95\textwidth]{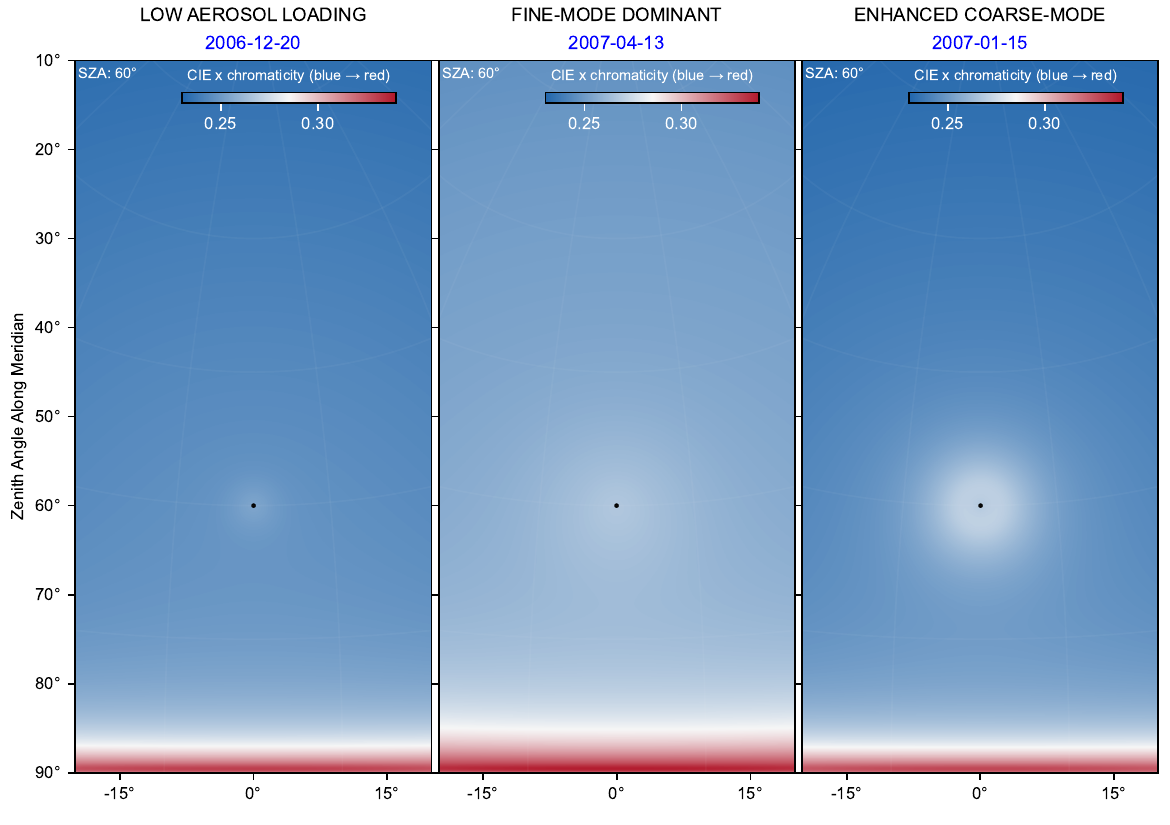}
\caption{CIE $x$–$y$ chromaticity renderings corresponding to the synthetic circumsolar sky images in Figure~\ref{fig:synth_sky}. By isolating color information independent of luminance, these maps distinguish spectral reddening associated with fine-mode aerosol scattering from largely achromatic brightening characteristic of coarse-mode forward scattering. An animated version of this figure, showing the chromaticity evolution with changing solar elevation, is available in the online journal.}
    %\end{interactive}
    \label{fig:synth_sky_xchroma}
\end{figure*}

\section[Physically Based Color Rendering of Circumsolar Sky Appearance]{Physically Based Color Rendering of \\ Circumsolar Sky Appearance}
\label{sec:synthColor}

The preceding sections show that aerosol optical properties retrieved from AERONET provide sufficient information to diagnose circumsolar sky brightness at angular scales relevant for coronal observations.  To showcase the additional information the AERONET data provides in this context, here we apply a wavelength-resolved radiative-transfer framework constrained by AERONET aerosol properties to generate synthetic sky images for the representative aerosol regimes introduced in Section~\ref{sec:daily_comparisons}. The objective is not to reproduce individual SBM observations, but to extend the validated aerosol--radiance relationships established in Sections~\ref{sec:aeronet_model} and~\ref{sec:statistics} into an observer-oriented visualization of circumsolar sky appearance.

Synthetic sky radiances are computed at the elevation of Mauna Loa across the visible spectrum using the \texttt{uvspec} radiative-transfer model with DISORT \citep {stamnes1988} as the multiple-scattering solver, as implemented in \texttt{libRadtran} \citep{Mayer2005,Emde2016}. The solar source spectrum is specified using the extraterrestrial spectral irradiance of \citet{gueymard2004}, sampled at 1~nm resolution. We adopt a standard tropical atmospheric constituent profile with molecular concentrations scaled by AERONET measurements. The aerosol stratification is assumed to follow an exponential vertical profile with an e-folding height of 2 km above Mauna Loa.  Calculations employ the pseudo-spherical approximation to account for atmospheric curvature at large solar zenith angles while retaining a plane-parallel treatment of multiple scattering within each layer. Aerosol optical properties are prescribed entirely by contemporaneous AERONET data products. Aerosol optical depth, single-scattering albedo, and phase functions are taken from direct-Sun and inversion retrievals at the standard AERONET wavelengths (440, 675, 870, and 1020~nm). These quantities are parameterized as smooth functions of wavelength to enable continuous spectral modeling while preserving normalization and forward-scattering behavior.  Using these inputs, a grid of scattering angles are computed in both the azimuthal and zenith directions away from the solar zenith angle.  We compute all of these as a function of solar zenith angle, so as to generate a synthetic sunrise time sequence.

For color rendering, the wavelength-resolved radiance fields, in units of $\mathrm{W\,m^{-2}\,sr^{-1}\,nm^{-1}}$, are first converted to standard CIE XYZ tristimulus values using the 1964 color matching functions for a $10^{\circ}$ observer \citep{CIE2019_CMF10deg}.  The XYZ values are then scaled by a linear exposure gain factor to obtain sufficient dynamic color range of the final image in the standard RGB color space. This gain is dimensionless and is selected to minimize color saturation across the rendered field of view. Importantly, to ensure valid comparisons of different conditions, this gain is applied uniquely as a function of solar zenith angle (SZA). Once scaled, the XYZ values are translated into linear RGB values using a D65 conversion matrix and then gamma corrected with standard methods.

Figure~\ref{fig:synth_sky} presents synthetic true-color circumsolar sky images generated for the three representative Mauna Loa aerosol regimes introduced in Section~\ref{sec:daily_comparisons} for a SZA of $60^{\circ}$. These renderings provide a visual manifestation of the scattering behavior quantified in earlier sections.  For real world examples, see \citet{Mims2003}.  Here, Rayleigh-like conditions exhibit deep blue skies and a nearly non-existent aureole around the sky; meanwhile, fine-mode aerosols yield broader aureoles as well as a brighter haze over much of the rendered FOV.  The latter is consistent with the increase of diffuse-to-direct irradiance seen in Figure~\ref{fig:compare3days}(h). In contrast, coarse-mode–dominated conditions generate intense forward-scattering lobes that substantially increase near-Sun brightness.

With the exception of the region near the horizon, any reddening of the circumsolar sky is difficult to isolate visually independent of absolute luminance. The two qualities are respectively linked to the total aerosol loading (AOD) and the related spectral slopes (both $\alpha$ and $\gamma$).  In Figure~\ref{fig:synth_sky_xchroma}, the CIE x chromaticity value ($x = X/(X + Y + Z)$), which represents the relative amount of red within the full color expression, is displayed across the rendered field-of-view.  This gives an independent assessment of reddening within the visual wavelength range of the sky.  Note that the colormap of Figure~\ref{fig:synth_sky_xchroma} is not true color in this case but instead reflects more or less values of red.  Clearly, the presence of coarse mode aerosols reddens the circumsolar region as consistent with its reduced spectral slopes. 

In addition to the static renderings, animated versions of Figures~\ref{fig:synth_sky} and~\ref{fig:synth_sky_xchroma} are provided in the online journal. The animations show the evolution of the synthetic circumsolar sky as solar elevation changes under fixed aerosol conditions, illustrating how apparent sky brightness and color can vary substantially even in the absence of aerosol variability.

These renderings highlight that visually apparent mild haze and an extended solar aureole do not necessarily imply poor infrared coronal sky conditions: in all three cases, the near-Sun sky brightness at infrared wavelengths remains $\lesssim 2\,\mu B_{\odot}$ per airmass. These conditions are generally considered good for coronal observations; and so, one should not misinterpret the presence of light haze and/or a pronounced solar aureole as definitively indicative of poor daytime sky conditions. 

%%%%%%%%%%%%%%%%%%%%%%%%%%%%%%%%%%%%%%%%%%%%%%%%%%%%%%%%%%%%%%%%%%%%%%%%
%%% SECTION 7
%%%%%%%%%%%%%%%%%%%%%%%%%%%%%%%%%%%%%%%%%%%%%%%%%%%%%%%%%%%%%%%%%%%%%%%%

\section{Discussion}
\label{sec:discussion}

In this study, we have compared complementary diagnostics useful for assessing daytime sky clarity for solar coronal imaging. The circumsolar region within $\sim\!2^{\circ}$ of the Sun's disk center is operationally critical yet the most difficult region to measure. For site surveys, daily assessments, and long term monitoring, solar observatories have utilized tools such as the Evans Solar Photometer \citep{evans1948} and the ATST Sky Brightness Monitor \citep{Lin2004PASP} to quantify the near-Sun radiance at different locations. Given the strong influence of coarse-mode aerosols, some site surveys have also considered dust counting measurements \citep{Hill2006} and related turbidity metrics. Measurements of sky brightness at Dome C, Antarctica, have been analyzed alongside Lidar-based atmospheric disturbance indices \citep{Haudemand2024}. \citet{Iglesias2023} reviewed mean particle concentrations at El Leoncito, Argentina, as part of their coronagraphic sky brightness assessment.

The current study benefits from co-temporal coronagraphic measurements (\textit{i.e.} SBM) of the near-Sun sky brightness and Sun--sky radiometric measurements (\textit{i.e.} AERONET) used for aerosol optical inversions. Here, we demonstrate both qualitative and quantitative consistency between SBM measurements and AERONET-derived inferences, and we provide a model-based method for inferring the near-Sun brightness based on almucantar sky radiance data and the associated inverted data products. The quantitative comparisons between these inferences and the SBM data provide confidence that either method can be used to quantify the near-Sun brightness spectrally and across multiple timescales.

Meanwhile, a range of other measures are also available to quantify atmospheric conditions, including monochromatic aerosol optical depth, the Ångström exponent, the diffuse-to-direct irradiance ratio, and/or coarse-mode aerosol volume concentration. As seen in Figure~\ref{fig:size_dist}, the coarse-mode aerosol volume concentration is perhaps the most predictive quantity for near-Sun brightness; however, the scatter remains significant enough to warrant caution, as fine-mode accumulation can also lead to increased near-Sun radiance. Likewise, the variables shown in Figure~\ref{fig:gammaStats} are not uniquely predictive of near-Sun brightness. As per Equation~\ref{eq:sky_lam}, the near-Sun radiance is an interplay of aerosol loading, single-scattering albedo, and the particle phase function. These further relate to particulate size and shape distributions as well as their complex, spectrally dependent indices of refraction.

As mentioned previously, our study has been limited to Mauna Loa Observatory, which is a notably pristine site for coronagraphic observations. While the results provide a strong case that Sun--sky radiometric data and coronagraphic measurements may be used jointly or separately to quantify the near-Sun radiance at angles $\lesssim \!2^{\circ}$, we have not fully investigated how different aerosol conditions and properties may skew the relevant uncertainties. Furthermore, we have used the AERONET data products with a focus on reproducing the near-Sun almucantar radiance curves and not the inherent accuracy of the retrieved quantities themselves. A number of AERONET related studies, notably \citet{Dubovik2000Accuracy} and \citet{Sinyuk2020}, quantify uncertainties in retrieved optical properties for diverse conditions but not necessarily as a function of scattering angle. Meanwhile, there is ongoing interest in assessing whether AERONET data can be used to determine the contribution of circumsolar total irradiance that is relevant to solar energy estimation \citep{DeVore2009, eissa2015, abreu2020}.

We also remind the reader of approximations used in this study. The exoatmospheric solar irradiance is assumed unidirectional, \textit{i.e.}, the Sun is considered as a point source. Center-to-limb variation of the solar radiance is accounted for only when reporting the disk-normalized sky brightness values; it is not considered within the almucantar radiance approximation of Equation~\ref{eq:sky_lam}. Furthermore, both the AERONET radiometric data and the SBM imaging data are averaged over fields of view of greater than one degree on sky, which is not explicitly treated in our analysis. Likewise, there are variations in airmass within individual data acquisitions of either instrument, due e.g., to field of view coverage and/or inherent sky non-uniformity during an almucantar scan. These and other error terms are recognized and should be improved upon in the future; however, they do not fundamentally impact the statistical correspondences and data representations presented here.

From a solar astronomy perspective, it is worth recognizing the additional confidence in site sky-clarity assessments that is provided by direct measurements such as those from the SBM. After all, Sun–sky photometers do not typically sample the strong forward peak of coarse-mode scattering. Nevertheless, for the case of Mauna Loa, we have demonstrated that these two approaches yield consistent estimates of near-Sun sky brightness. At the same time, SBM-like coronagraphic measurements are not immune to systematic errors or incomplete instrumental characterization. Aerosol optical inversions therefore provide a viable and independent means of constraining the near-Sun radiance field, provided that the uncertainties in the derived forward-scattering phase function are properly understood. More broadly, Sun–sky photometry expands site characterization to a regime in which physics-based atmospheric modeling becomes tenable, enabling this study to produce realistic, physically motivated true-color renderings of the Mauna Loa sky under varying aerosol conditions.

While there is a rich literature, in particular using AERONET data, pertaining to the microphysics of the atmosphere and its dynamical evolution across the globe, this information has to date been underutilized for solar coronal observing site assessment. For existing coronagraphic facilities like the US National Science Foundation's Daniel K Inouye Solar Telescope \citep[DKIST;][]{rimmele2020} and future facilities under study, like the Coronal Solar Magnetism Observatory \citep[COSMO;][]{tomczyk2016}, the inclusion of Sun--sky radiometric metrology should be strongly considered to increase the value and completeness of both short and long-term monitoring of daytime circumsolar conditions. Meanwhile, the methodology outlined in this work can and should be extended to assess other potential coronal sites with existing sun-sky photometry data.

\vspace{3mm}

\begin{acknowledgments}
We thank the Mauna Loa AERONET principal investigators, Pawan Gupta and Elena Lind, for their effort in establishing and maintaining the Mauna Loa AERONET site.  We also extend our appreciation to Steve Tomczyk and David Elmore for deploying the SBM to Mauna Loa and calibrating the data, as well as Haosheng Lin for his development of the SBM.  Sarah Jaeggli, Ben Berkey, Joan Burkepile provided useful feedback and information in support of this work.  Thanks also to Africa Barreto and Pablo Gonzalez for helpful discussions regarding AERONET. This material is based upon work supported by the National Center for Atmospheric Research, which is a major facility sponsored by the National Science Foundation under Cooperative Agreement No. 1852977.  The National Solar Observatory (NSO) is managed by the Association of Universities for Research in Astronomy, Inc., and is funded by the National Science Foundation. Any opinions, findings and conclusions or recommendations expressed in this publication are those of the author(s) and do not necessarily reflect the views of the National Science Foundation or the Association of Universities for Research in Astronomy, Inc. We acknowledge the use of ChatGPT (OpenAI) for language editing and text drafting support under the careful scrutiny of the authors.
\end{acknowledgments}

%\facility{DKIST}

\appendix
\twocolumngrid

\section{Distributions of Retrieved Closure Terms}\label{sec:appendix1}

\begin{figure}
    \centering
    \includegraphics[width=0.95\columnwidth]{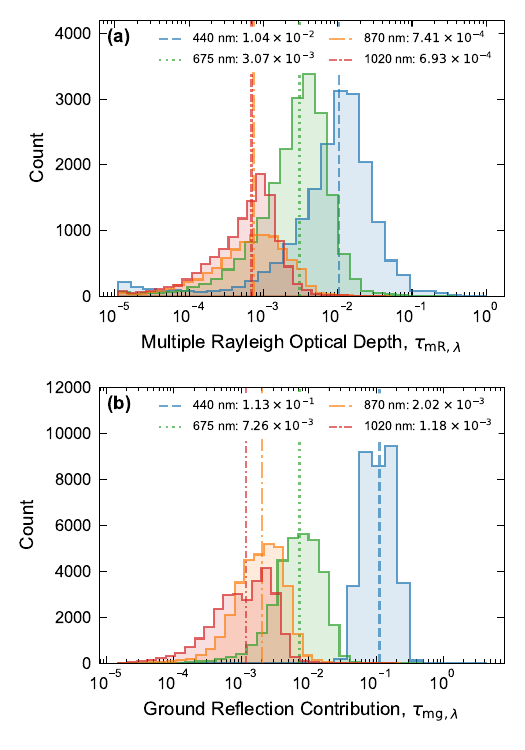}
    \caption{
Distributions of the fitted multiple-scattering optical depth terms derived from the analytical modeling fitting of the almucantar sky radiance data at four wavelengths (440, 675, 870, and 1020\,nm). (a) The multiple Rayleigh optical depth component, $\tau_{\mathrm{mR},\lambda}$ and 
(b) Ground-reflection contribution, $\tau_{\mathrm{mg},\lambda}$. Histograms show the distribution of retrieved values across all valid almucantar inversions; vertical dashed lines denote the median for each wavelength, listed in the legend.}
    \label{fig:multiple_scatter}
\end{figure}

Figure~\ref{fig:multiple_scatter} shows the distributions of retrieved almucantar closure terms, $\tau_{\mathrm{mR},\lambda}$ and $\tau_{\mathrm{mg},\lambda}$. Both exhibit distinct, wavelength-dependent modal peaks.  \citet{Gueymard2001} describes $\tau_{\mathrm{mR},\lambda}$ as a ``fictitious'' optical thickness that enters as an additional Rayleigh-like component arising from multiple scattering. More precisely, it serves as a heuristic parameterization that approximates second-order scattering effects involving both molecular and aerosol scattering. The characteristic single-scattering Rayleigh optical depths at Mauna Loa, $\tau_{\mathrm{R},\lambda}$, are 0.163, 0.028, 0.010, and 0.005 at 440, 675, 870, and 1020\,nm, respectively. The squares of these values—$2.7\times10^{-2}$, $8\times10^{-4}$, $1\times10^{-4}$, and $2.5\times10^{-5}$—are comparable to the modal values in panel (a), consistent with the expected second-order scaling of multiple scattering. The extended tails toward smaller values likely reflect increased retrieval uncertainty near the measurement noise floor. Meanwhile, $\tau_{\mathrm{mg},\lambda}$ represents a backscattering contribution from ground reflections. \citet{Box1979} estimates this term as $\tau_{\mathrm{mg},\lambda} = \rho_{\mathrm{g},\lambda} \left(\tau_{\mathrm{R},\lambda} + \tau_{\mathrm{mR},\lambda}\right)$, where $\rho_{\mathrm{g},\lambda}$ is the spectral ground albedo. The distributions in panel (b) are broadly consistent with this scaling, particularly given the low surface albedo expected for the volcanic terrain of Mauna Loa.

\bibliography{main}{}
\bibliographystyle{aasjournalv7}

\end{document}